\documentclass[twocolumn,showpacs,preprintnumbers,amsmath,amssymb]{revtex4}
\usepackage{graphicx}
\usepackage{epsfig}
\usepackage{dcolumn}
\usepackage{bm}
\usepackage{epstopdf}
\begin{document}
\preprint{LPT-Orsay : 11-92}
\title{Melting in Monolayers : Hexatic and Fluid Phases.}
\author{Martial MAZARS}
 \email{Martial.Mazars@th.u-psud.fr}
\affiliation{%
Laboratoire de Physique Th\'eorique (UMR 8627), Universit\'e de Paris-Sud and CNRS, B\^atiment 210, 91405 Orsay Cedex, FRANCE }%
\date{\today}
\begin{abstract}
There are strong evidences that the melting in two dimensions depends crucially on the form and range of the interaction potentials between particles. We study with Monte Carlo simulations the phase diagram and the melting of a monolayer of point-particles interacting with repulsive Inverse Power Law Interactions, $V(r)=Q^2(\sigma/r)^n$ where $n$ can take any real positive value ($n$-OCP monolayer). As $n$ is varied from $0$ to $\infty$ (Hard Disks), including Coulomb ($n=1$) and Dipolar ($n=3$), melting occurs with different mechanisms and the overall picture permits to understand the diversity of mechanisms found experimentally or in computer simulations for 2D melting. The empirical transition curves for $n\leq 3$ and the excellent qualitative and semi-quantitative agreements with the KTHNY theory found for the melting of $n$-OCP monolayers with $n\leq 3$ are the main results of the present work.  
\end{abstract}
\pacs{64.70.D-, 61.20.Ja, 64.60.De} 
\maketitle

\section{Introduction}
The crystalline phases in two dimensional systems are characterised by long-range orientational order and, because of the importance of fluctuations \cite{Mermin:68,Jancovici:67}, a quasi-long range positional order rather than a long ranged order.  Several experimental systems, as electrons at surface of liquid Helium \cite{Monarkha:book:03,Grimes:79,Mehrotra:82,Gallet:82} or in MOSFETs transitors (Wigner crystals) \cite{Monarkha:book:03,Platzman:74,Muto:99,He:03,Clark:09}, dusty plasma \cite{Nosenko:09}, colloids confined in slabs or at interfaces \cite{Murray:87,Keim:07,Marcus:97,Dullens:04,Zheng:06,Peng:11}, Abrikosov vortex in 2D supraconductors \cite{Iaconis:10,Saiki:11}, rubidium atoms trapped in 2D optical traps \cite{Hadzibabic:06,Choi:12}, atoms adsorbed at liquid or solid surfaces \cite{Negulyaev:09}, may be considered as 2D systems. Most of these systems exhibit phase transitions between their crystalline and disordered (hexatic or isotropic) fluid-like phases. To describe 2D meltings, a variety of mechanisms has been proposed \cite{Strandburg:88}, among them, there are the grain-boundary induced melting \cite{Chui:82} where the Solid-Fluid transition is described as a first order phase transition between the solid and the fluid phases, and the KTHNY theory (after Kosterlitz, Thouless, Halperin, Nelson and Young) \cite{Nelson:book:83,Nelson:78,Nelson:79,Young:79,Zippelius:80,Kosterlitz:74,Saito:82,Saito:83}, with an intermediate hexatic phase and two continuous phase transitions (Solid/Hexatic and Hexatic/Fluid). The melting in two dimensions may also be described in the vicinity of the isostructural critical points \cite{Bladon:95,Marcus:96,Chou:96}. The isostructural transition describes a Solid-Solid transition in which the two Solid have the same symmetry but a different lattice spacing. The hexatic phases in systems with isostructural transition occur when the triple point (Fluid-Solid-Solid) is close enough to the isostructural critical point (Solid-Solid) \cite{Bladon:95,Chou:96}.\\
From experimental and numerical simulations results, it is clear that the underlying mechanisms for 2D meltings depend on the geometry of the systems \cite{Peng:11,Gribova:11} and interactions between constituants ; in particular, the melting depends strongly on the range and form of the interaction \cite{Nosenko:09,Prestipino:11,Clark:09,Keim:07,Bernard:11,Jaster:99,Bladon:95,Marcus:96,Chou:96,Sengupta:00,Pronk:04,Mak:06,Wierschem:11,Tobochnik:82,Shiba:09,Asenjo:11,Gribova:11}.\\
In this paper, we report a preliminary study of the melting in monolayers of point particles that interact with repulsive Inverse Power Law (IPL) potentials. The repulsive Inverse Power Law interactions are
\begin{equation}
\label{IPLpot}
\displaystyle V(r)=Q^2\left(\frac{\sigma}{r}\right)^n
\end{equation}
where $Q^2$ is the energy scale, $\sigma$ the length scale and $r$ the distance between the particles. In this class of potential, the power $n$ can be considered as a parameter and it shall take any real positive value in the following \cite{Mazars:11,Mazars:10,Hoover:71,Khrapak:11,Groh:01,Pedersen:10,Coslovich:08,Agrawal:95}. Although $n$ is considered in this paper as a continuous parameter, some integer values of $n$ correspond to physical interactions in experimental and real systems, as for instance : Coulomb interactions ($n=1$) \cite{Monarkha:book:03,Grimes:79,Mehrotra:82,Gallet:82,Platzman:74,Muto:99,He:03,Clark:09,Nosenko:09,Zheng:06,Lidmar:97}, dipolar interactions ($n=3$) \cite{Keim:07,Lin:06}, the short range repulsion in the Lennard-Jones potential ($n=12$) \cite{Wierschem:11,Tobochnik:82,Shiba:09,Asenjo:11,Gribova:11} and also the hard disk interaction in the limit $n\rightarrow\infty$ (then $\sigma$ is the diameter of the disks)\cite{Bernard:11,Jaster:99,Sengupta:00,Pronk:04,Mak:06,Isobe:10,Isobe:11}. For all these interactions the melting in two dimensions has been studied experimentally and numerically ; different mechanisms have been found including the grain boundary induced melting \cite{Chui:82,Strandburg:88}, the KTHNY theory \cite{Nelson:book:83,Nelson:78,Nelson:79,Zippelius:80,Kosterlitz:74,Strandburg:88} and also, on hard disks systems, a first order transition from the isotropic fluid to the hexatic phase followed by a continuous transition from the hexatic phase to the 2D-hexagonal crystal \cite{Bernard:11}. Since the interaction between the particles in all these systems belongs to the class of repulsive Inverse Power Law potentials, by considering the power $n$ as a continuous parameter we are able to define a model on which the crossover between all the mechanisms for the two dimensional melting can be studied.\\
It is worthwhile to outline also that repulsive Inverse Power Law Interactions are used as repulsive reference potentials in dense liquids to study the thermodynamic and dynamical properties \cite{Pedersen:10,Coslovich:08,Agrawal:95}. In these models, IPL potentials are effective interactions and the exponent $n$ is computed from the scaling properties of the diffusion coefficients \cite{Coslovich:08} or by using the virial theorem \cite{Pedersen:10} ; the values of $n$ computed likewise are in general greater than 10. Repulsive Inverse Power Law interactions are also used in studies on glass transitions \cite{Michele:04,Coslovich:08,Berthier:10,Casalini:06}.\\
The purpose of this paper is to define the $n$-OCP monolayer model and to give a first overview of its properties and phase diagram, most of the results reported in the present work are obtained for long ranged and quasi long ranged interactions ($n\leq 2 - 3$).\\
The paper is organised as follows. In section II, we describe the numerical methods and the model system used in this study and we give computational details on the Monte Carlo simulations. Based on the results obtained by simulations, the empirical transition curves as function of $n$ are given in section III. In section IV, we report the results of Monte Carlo simulations done in the present work. In this section, a particular attention is paid to the Kosterlitz-Thouless essential singularity \cite{Kosterlitz:74} near the Fluid/Hexatic transition, since this singularity is a clear signature of the KTHNY theory \cite{Nelson:book:83}. A discussion of the results reported in the paper is done in section V.\\
For completeness, in an Appendix we give the Ewald sums for repulsive Inverse Power Law interactions in monolayers, a complete analytical derivation can be found in refs.\cite{Mazars:10,Mazars:11}.
\section{Methods and computational details}
To explore the phase diagram of a system of point particles with pairwise repulsive inverse power law interactions we use Monte Carlo simulations. The computations are done in the canonical ensemble with a variable shape of the simulation box, but at a constant surface area $A$ \cite{Mazars:08,Weis:01}. The Ewald method for inverse power law interactions derived in refs.\cite{Mazars:10,Mazars:11} is used in all computations. For long and intermediate ranged potentials ($n\leq7$) the use of Ewald methods is necessary to avoid bias in computations. If the interactions are very long ranged ($n\leq 2$), a neutralizing background is used to obtain convergent sums as in the OCP model \cite{Totsuji:78} ; in the following, the systems of point particles with Inverse Power Law potential interactions Eq.(\ref{IPLpot}) with a neutralizing background when convergence is needed \cite{Mazars:10,Mazars:11,Smith:08} are called $n$-OCP systems.\\
For short ranged interactions (repulsive soft spheres - $n > 10$) truncations of the interactions can be done safely, however, in this study we use Ewald methods also for these values of $n$ to avoid any type of discontinuity that could be introduced by changing the method used to compute energies. In the appendix, one would find the energy of the monolayer computed with the Ewald method.\\ 
All thermodynamic points are obtained on systems with $N=4096$ particles, equilibration is done by making $2\times 10^5$ MC-cycles and the number of MC-cycles to compute averages are between $2\times 10^5$ and $2.3\times 10^6$. With the code written for this study, the cpu-time for one MC-cycle is about one or two seconds ; however, despite this quite good efficiency, the computation time needed to obtain one thermodynamic point is about one or two months. Improvements of the source code with mesh methods (PME, P3M, etc.) \cite{Mazars:11} can not be done easily for non integer value of the power $n$, especially for the grid points fractional charge assignment step \cite{Mazars:11}.\\
The reduced units are defined as follows : density $\rho^*=\rho\sigma^2$ ; coupling $\tilde{\Gamma}=Q^2/k_B T=Q^{*2}$, all computations are done with $Q=14$. For this system, the only relevant parameter is the coupling constant $\Gamma= (\pi\rho \sigma^2)^{n/2} \tilde{\Gamma}$. For notational convenience, the asterisks will be dropped in the remaining of the paper. \\
Bond orientational order parameters (BOOP), their histograms and their correlations are computed by using Voronoi constructions. It is worthwhile to note that Voronoi (or Delaunay) constructions to compute order parameters and a variable shape of the simulation box are necessary to study fluid-solid phase transitions in two dimensions to avoid systematic bias \cite{Frenkel:12}. Voronoi construction permits to identify exactly the neighbors of all particles and variation of the shape of the simulation box is necessary to avoid irrelevant hysteresis loops induced by some inadequacies between the geometry of the box and the geometry of the cristal or hexatic phases, even for large and very large systems.\\ 
The finite size analysis presented in section IV is achieved by dividing the system in smaller subsystems in a similar way as it is done in ref.\cite{Weber:95}, the fourth-order cumulants  \cite{Privman:book:90} are defined as 
\begin{equation}
\label{Cumulant}
\displaystyle U_6(N)=1-\frac{<\phi_6^4>_N}{3<\phi_6^2>_N^2}
\end{equation}
with $N$ the average number of particles in subsystems. A more detailed study using different system sizes and the multiple histogram reweighting method  \cite{Ferrenberg:88,Newman:book:99} is on-going. Some results obtained with the multiple histogram method are given in section IV for systems with 4096 and 2025 particles and for $n=1.65$ and $\rho=0.109$.\\ 
To obtain measurements of the correlation lengths and exponents of the algebraic decay of correlations in hexatic and fluid phases, the tails of correlation functions ({\it i.e.\/} for $s> 10\mbox{ }\sigma$) are fitted as follows. The tails of bond orientational correlation functions in the Fluid phase are represented as 
\begin{equation}
\label{Fit_g6_fluid}
\begin{array}{ll}
\displaystyle g_6(s) &\displaystyle =<\phi_6>_g^2+G_6^{(F)}e^{-s/\xi_+(T)}\\
&\\
&\displaystyle +G_6^{(H)}e^{-s/\xi_6(T)}\mbox{ J}_0\left(2\pi\frac{s}{\lambda_6}\right)
\end{array}
\end{equation}
and, in Hexatic and Solid phases, as
\begin{equation}
\label{Fit_g6_hexatic}
\displaystyle g_6(s)= <\phi_6>_g^2+H_6\left(\frac{\sigma}{s}\right)^{\eta_6(T)}\mbox{ J}_0\left(2\pi\frac{s}{\lambda_6}\right)
\end{equation}
with $\mbox{ J}_0(x)$ the Bessel function. Similar forms are used for center to center correlation functions, with parameter $\xi(T)$, $\eta(T)$ and $\lambda$, as 
\begin{equation}
\label{Fit_gFH_fluid}
\displaystyle g(s)= 1+G_0\mbox{ }e^{-s/\xi(T)}\mbox{ J}_0\left(2\pi\frac{s}{\lambda}\right)
\end{equation}
in the Fluid and Hexatic phases, and as 
\begin{equation}
\label{Fit_gS_fluid}
\displaystyle g(s)= 1+G_S\left(\frac{\sigma}{s}\right)^{\eta(T)}\mbox{ J}_0\left(2\pi\frac{s}{\lambda}\right)
\end{equation}
in the Solid phase.\\
The center to center correlation functions $g_{pq}(s)$ for pair of particles with $p$ neighbors and $q$ neighbors are computed from the Voronoi constructions and normalised to 1 as $s\rightarrow\infty$. In Solid and Hexatic phases, since the particles with a number of neighbors different from 6 are quite rare, large statistical fluctuations in the $g_{pq}(s)$ functions are found for $s>10\sigma$ and extremely strong correlations are found for $s\sim 1/\sqrt{\pi\rho}$.\\
The Kosterlitz-Thouless essential singularity near transitions, for a system with $N$ particles, can be represented as \cite{Kosterlitz:74,Nelson:79,Privman:book:90}
\begin{equation}
\label{KT_sing}
\displaystyle X(\gamma;N)=X_{0}(N)\exp\left(\frac{a(N)}{\mid\gamma-\gamma_{c}(N)\mid^\nu}\right)
\end{equation}
with $\gamma_{c}(N)=1/\Gamma_c(N)$.\\
For MC computations done with $n$, $\rho$ and $Q$ fixed, one can replace $\gamma_{c}(N)$ by $T_c(N)$. All quantities exhibiting a KT-singularity (correlation lengths and quantities related to them) are fitted with Eq.(\ref{KT_sing}) to obtain estimates of $T_c(N)$ or $\Gamma_c(N)$ \cite{Keim:07}. For a given set of computations (at fixed $n$ and $\rho$, or at fixed $\rho$ and $T$), two temperatures or coupling constants are found ; we identify one with the Fluid/Hexatic transition ($T_{F/H}$ or $\Gamma_{F/H}$) and the other with the Hexatic/Solid transition ($T_{H/S}$ or $\Gamma_{H/S}$). We locate the Fluid/Hexatic transitions from the KT-singularities found for : the finite-size analysis of $<\phi_6>$ and $\chi_6$ ($N\rightarrow\infty$ - cf. Inset FIG.\ref{Fig4}(b)), $\xi(T)$ and $\xi_6(T)$ ; and from $\xi_{+}(T)$ for the Hexatic/Solid transitions. The values found for $\Gamma_c(n)$ by doing these analysis are reported on TABLE \ref{Table1} and the value of $\rho_c(n)$ deduced from $\Gamma_c(n)$ at $T=1$ is also given on TABLE \ref{Table1} and represented as symbols denoted MC on FIG.\ref{Fig1}(a). As explained in the next section, the values of $\Gamma_c(n)$ permit also to construct empirical transition curves, for the two transitions, as function of the power $n$ of the IPL potentials.
\section{The empirical transition curves as function of $n$}
The main purposes of this paper are to obtain a first overview of the phase diagram of the monolayer $n$-OCP model as function of the parameter $n$ and to locate approximately the critical or multicritical points in the $(n,\rho)$ plane or $(n,\Gamma)$ plane. The coupling constant $\Gamma$ is the only relevant parameter for the $n$-OCP system. Previous experimental and numerical studies on Coulomb, dipolar and hard disk systems show that the value of $\Gamma_c$ at transition depend on the interaction. Therefore, we may represent the phase diagram of the monolayer $n$-OCP model in the $(n,\Gamma)$ plane with the help of curves $\Gamma_c(n)$ or equivalently, in the $(n,\rho)$ plane as curves 
\begin{table*}
\begin{tabular}{ccc  | ccc| c cc}
\hline
\hline
                            &                       &                                 &                              &                        &                              &                                  &                                          &\\
\multicolumn{3}{c|}{Fluid/Hexatic}  & \multicolumn{3}{c|}{Hexatic/Solid}  & & & \\
                            &                       &                                 &                              &                        &                              &                                  &                                          &\\
\hline
$n$                      & $\rho_c(n)$ & $\Gamma_c(n)$   & $n$                     & $\rho_c(n)$  & $\Gamma_c(n)$ & Transitions            &  References                   & Methods \\
\hline
                            &                       &                                 &                              &                        &                              &                                  &                                          &\\
HD                      &                       &                                 &HD                        &                      &                            &                            &                             &\\
($n\equiv\infty$) &   0.891-0.912   & $\infty$        &($n\equiv\infty$) &    0.917         &        $\infty$         &  $1^{st}/2^{nd}$  & \cite{Bernard:11} & Simulations (MC) \\
                    &                     &                              &                           &                      &                            &                            &                             &\\
\hline
                             &                                   &                                 &                                 &                                        &                                  &                  &                     &\\
$0.66\pm 0.09$ & $1.35/\pi$                &   $216.4\pm 9.5$  &  $0.71\pm 0.05 $  &  $1.35/\pi$                    &  $217.9 \pm 6.0$  & KTHNY    & This work   & Monte Carlo (A)\\
0.71      	          & $0.312\pm 0.018$  &   $194.7\pm 1.2$  & 0.71           	        &   $0.388\pm 0.011$   &  $210.2 \pm 1.0 $   & KTHNY    & This work & Monte Carlo  \\
$0.72\pm 0.02$ &  0.3                            &   $191.9\pm 2.9$  &  $0.86 \pm 0.02$  &  0.3                                &   $191.1 \pm 5.0$  & KTHNY    & This work  & Monte Carlo (B) \\
1                          & 0.123 - 0.192           &   $137\pm 15$        & 1                             &  0.123 - 0.192   	    &  $137\pm 15$    &        &\cite{Grimes:79,Monarkha:book:03,Mehrotra:82,Gallet:82}& Experiments \\
1			&  0.119                        &  120                         & 1                             &  0.163                            & 140                    &              & \cite{Clark:09} & Quantum Monte Carlo\\
1			&  0.123                        &  122                         & 1                             &  0.127                            & 124                   &  KTHNY          & \cite{He:03} & Monte Carlo\\
1.48		          & $0.132\pm 0.016$  &   $102.3\pm 0.5$  & 1.48                        & $0.152\pm 0.025$     &  $113.5 \pm 1.0 $ & KTHNY       &This work &  Monte Carlo\\
1.65 	                   & $0.128\pm 0.010$   &  $92.9\pm 0.3$     & 1.65                        &  $0.138\pm 0.028$    &  $98.3\pm 1.0 $      & KTHNY     &This work &  Monte Carlo\\
3$^\dagger$      & 0.141	                     &	  57.5		 & 3$^\dagger$         &  0.147                          &  61.5                          & KTHNY     &\cite{Keim:07} & Experiments\\
3$^\ddag$ 	 & [$1/2\sqrt{3}$]             &   $69.1\pm 0.5$    & 3$^\ddag$          &  [$1/2\sqrt{3}$]             & $ 72.0\pm 1.0$        & KTHNY      & \cite{Lin:06} &  Molecular Dynamics\\
3 	                    & $0.159\pm 0.008$   &   $69.3\pm 0.4$    & 3                             &  $0.161\pm 0.019$   & $ 70.7\pm 1.0$        & KTHNY      &This work &  Monte Carlo\\
                     &                     &                              &                           &                      &                            &                            &                             &\\
\hline
\hline
\multicolumn{6}{c}{Fluid/Solid} & & & \\
\hline
                     &        $n$         &                              &          $\rho_c(n)$              &          &  $\Gamma_c(n)$         &                           &                  &\\
\hline
                     &                     &                              &                           &                      &                            &                            &                             &\\
                     &      7.30-7.40  &                              &          0.3                   &          &  $157.4-157.9$         & $1^{st}$ order    & This work & Monte Carlo (B)\\ 
                     &   12.75-12.95 &                              &          $1.35/\pi$        &          &   $1328-1368$          & $1^{st}$ order    & This work &  Monte Carlo (A) \\ 
                     &                     &                              &                           &                      &                            &                            &                             &\\
\hline
\hline
\end{tabular}
$^\dagger$See text ; $^\ddag$ The reduced units used in ref.\cite{Lin:06} are equivalent to $Q^2=1$ and the computations are done at the density indicated in the table, this density  does not correspond to the critical density at $T=1$ as in the other lines of the table.
\caption{\label{Table1} Estimations of densities and $n$-OCP coupling constants for the Fluid/Hexatic (F/H) and Hexatic/Solid (H/S) transitions for several values of $n$ at $T=1.0$. Data estimated in this work are extracted from the behaviour of the bond orientational correlation function $g_6(s)$ computed with Monte Carlo simulations and from the Kosterlitz-Thouless essential singularities. These data give the empirical transition curves as in Eq.(\ref{Fit_gamma}) and for $n\leq 3$ with : $\Gamma_{0}^{(F/H)}\simeq -0.84$ ; $\Gamma_{1}^{(F/H)}\simeq -1.42$ ; $\Gamma_{2}^{(F/H)}\simeq -0.16$ and $\Gamma_{0}^{(H/S)}\simeq -0.74$ ; $\Gamma_{1}^{(H/S)}\simeq -1.45$ ; $\Gamma_{2}^{(H/S)}\simeq -0.24$.}
\end{table*}
\begin{figure}
\centerline{\includegraphics[width=3.3in]{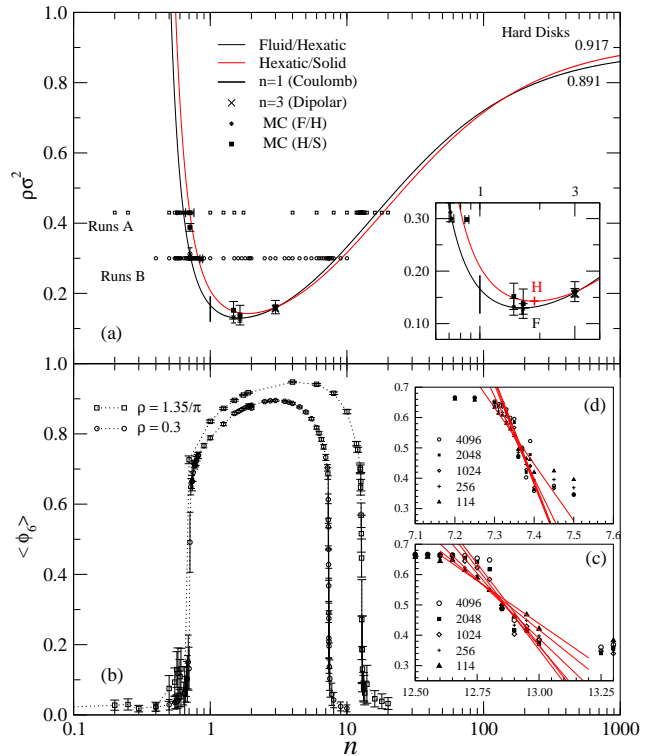}}
\caption{{\bf }(Color online) Phase diagram of monolayers ($T=1.0$ and $Q=14$). (a) Representation of the empirical transition curves for the Fluid/Hexatic and Hexatic/Solid transitions (Eq.(\ref{Fit_gamma})).  Inset : enlargement of the region $0.6\leq n\leq 3$,  locations of $F$ and $H$ are respectively $n_F\simeq 1.63$ and $\rho_F\sigma^2\simeq 0.130$ and $n_H\simeq 1.88$ and $\rho_H\sigma^2\simeq 0.143$. Runs A are MC computations done at $\pi \rho\sigma^2 =1.35$ and Runs B at $\rho\sigma^2=0.3$. (b) Bond orientational order parameters $<\phi_6>$ for Runs A and B. (c) Binder cumulants for Runs A near the F/S transition. The red lines are obtained by straight line fit of the cumulant in the region of transition, they are crossing for $n\sim 12.85-12.90$. (d) Same as (c), but for Runs B ; red lines cross around $n=7.35$.}
\label{Fig1}
\end{figure}   
\begin{figure}[ht]
\centerline{(a)\includegraphics[width=3.4in]{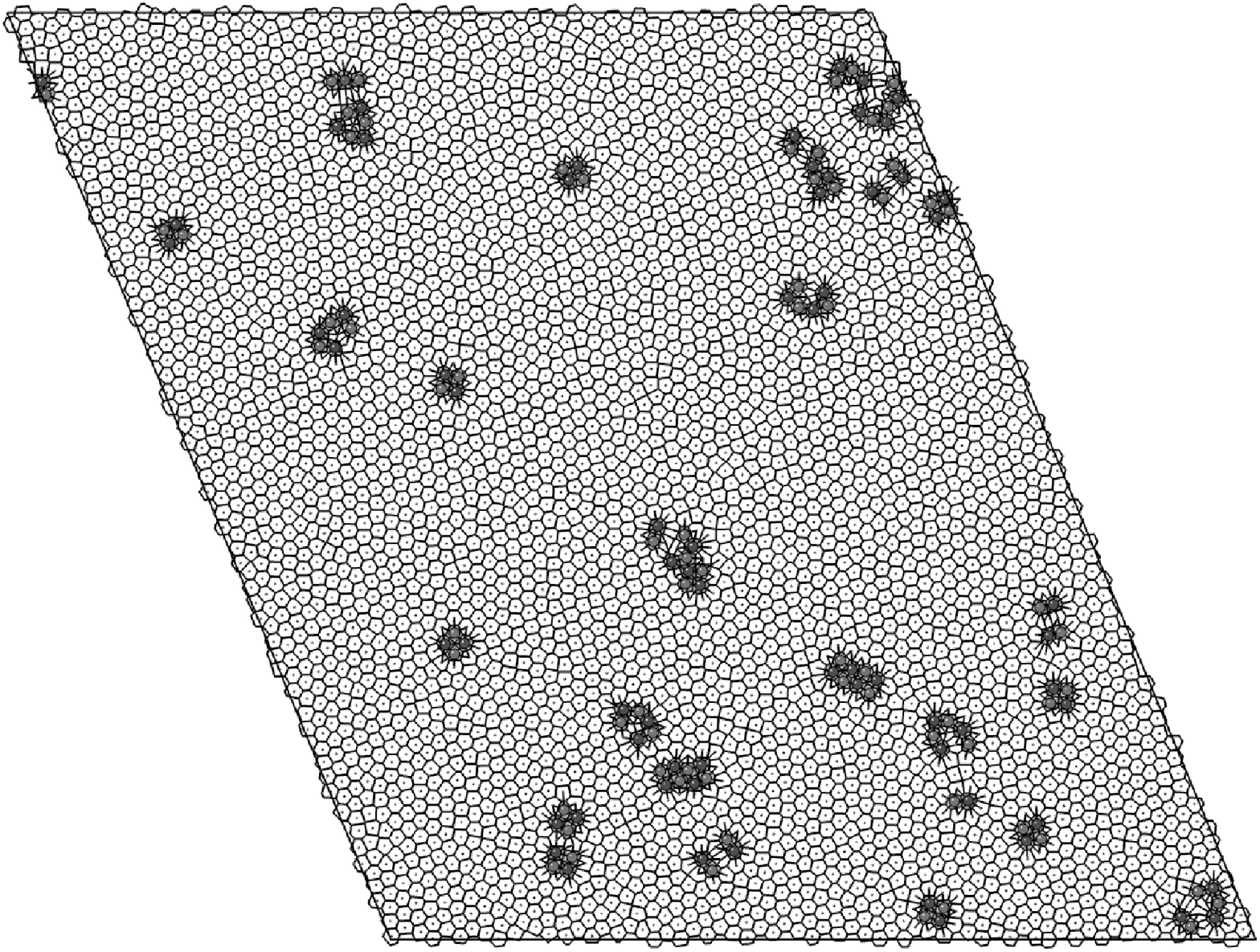}}
\centerline{(b)\includegraphics[width=3.4in]{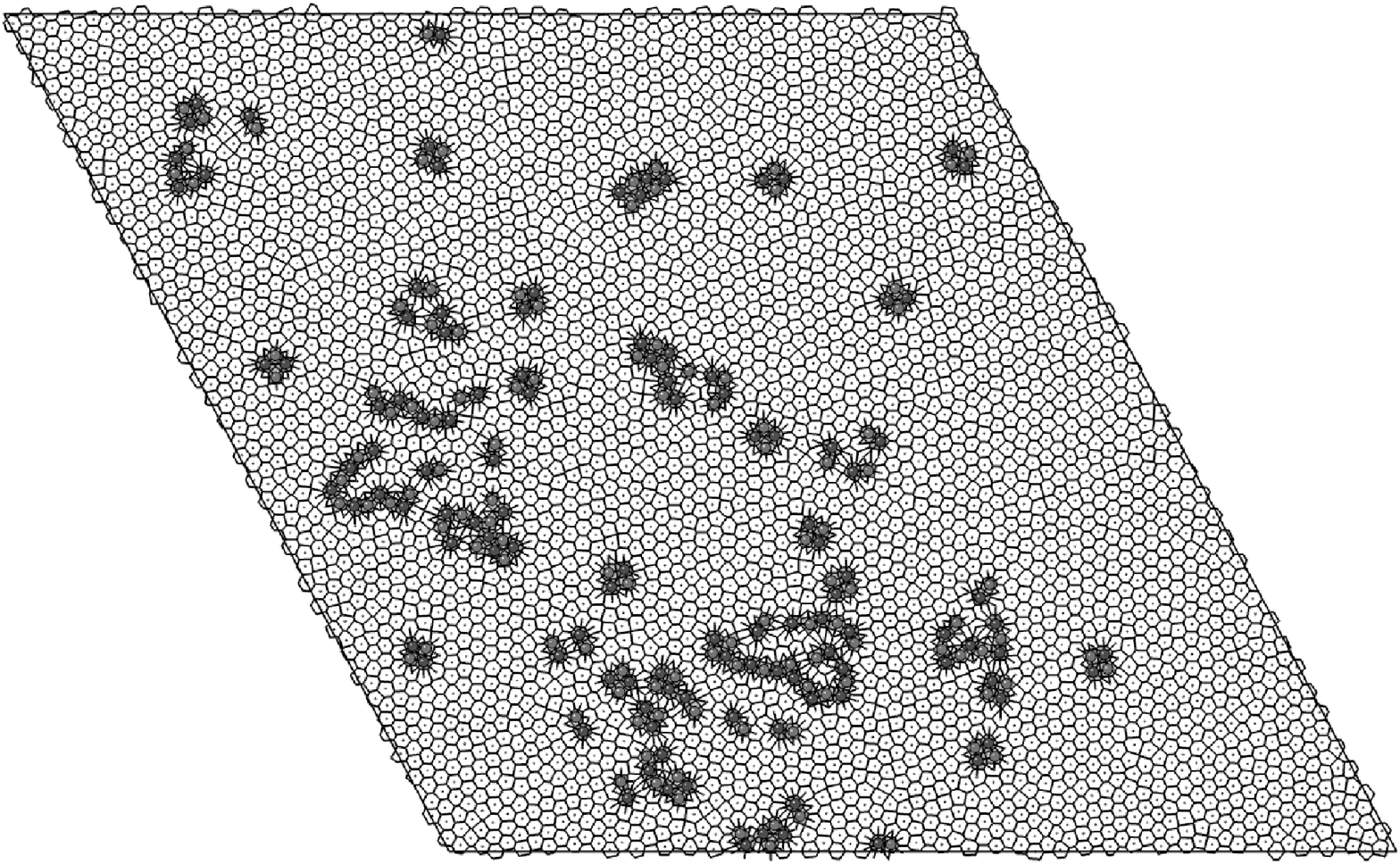}}
\caption{{\bf }(Color online) Snapshots of monolayer with Voronoi and Delaunay constructions near the Fluid/Hexatic transition. (a) The parameters are $n=0.71$,  $\rho=0.3$ and $\Gamma = 191.7$ ; for this thermodynamical state, the MC average of the  Bond orientational order parameters is $<\phi_6> =0.67\pm 0.03$. (b) $n=7.30$, $\rho=0.3$, $\Gamma=157.9$ and $<\phi_6> =0.61\pm 0.07$. White cells have 6 sides ; green : 5 sides ; red : 7 sides ; yellow : 4 sides  and blue : 8 or more sides.}
\label{Fig2}
\end{figure}
\begin{equation}
\label{MeanField}
\displaystyle \rho_c(n)=\frac{1}{\pi\sigma^2}\left(\frac{\Gamma_c(n) k_B T}{Q^2}\right)^{2/n}=\frac{1}{\pi\sigma^2}\left(\frac{\Gamma_c(n)}{\tilde{\Gamma}}\right)^{2/n}
\end{equation}
At present, we are not able to describe a theory that will give an analytical derivation of $\Gamma_c$ or $\rho_c$ as function of $n$. However, the numerical and the experimental results obtained is previous works \cite{Clark:09,Monarkha:book:03,He:03,Nosenko:09,Keim:07,Bernard:11} and the Monte-Carlo computations done in the present work allow us to give an empirical representation of the $\Gamma_c(n)$ and $\rho_c(n)$ curves.\\
On TABLE \ref{Table1}, we report some numerical evaluations of $\Gamma_c(n)$ and $\rho_c(n)$ for several systems including : Coulomb interactions ($n=1$) \cite{Clark:09,Monarkha:book:03,He:03,Nosenko:09}, dipolar interactions between superparamagnetic colloids ($n=3$) \cite{Keim:07,Lin:06}, simulations on Hard Disks systems ($n\rightarrow\infty$) \cite{Bernard:11} and MC computations done in the present work. These data are quite well represented by choosing $\Gamma_c(n)$ as
\begin{equation}
\label{Fit_gamma}
\displaystyle \ln\frac{\Gamma_c(n)}{\tilde{\Gamma}}=\Gamma_0+\Gamma_1\ln n+\Gamma_2(\ln n)^2+\Gamma_{\infty}n
\end{equation}
We assume the same dependence of $\Gamma_c$ on $n$ for both Fluid/Hexatic (F/H) and Hexatic/Solid (H/S) transitions but with different value of the $\Gamma_i$-parameters for each transition. For both transitions, $\Gamma_{\infty}$ may be obtained from the numerical computations done recently by Bernard and Krauth on hard disks systems \cite{Bernard:11}. The asymptotic behavior of Eq.(\ref{Fit_gamma})
\begin{equation}
\label{Gamma_infinity}
\displaystyle 2\lim_{n\rightarrow\infty}\left(\frac{\ln \Gamma_c(n)/\tilde{\Gamma}}{n}\right)=\ln\left(\pi \rho_c(\infty) \sigma^2\right)
\end{equation}
\begin{figure}[ht]
\centerline{\includegraphics[width=3.3in]{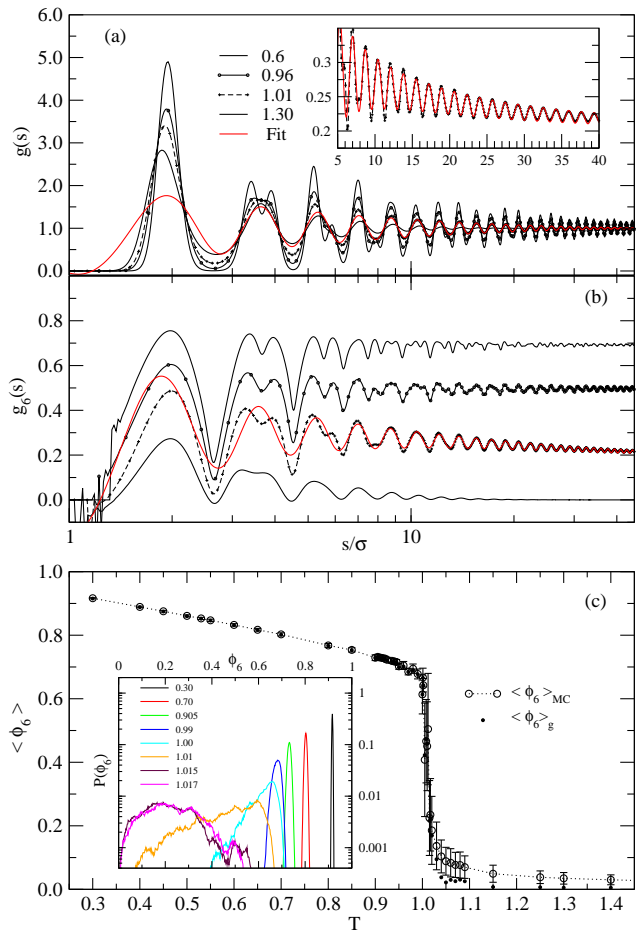}}
\vspace{-0.1in}
\caption{{\bf }(Color online) Dependence of the structure with temperature for $n=0.71$ and $\rho=0.3$. (a) Center to center correlation functions. The fit of the tail is represented for $T=1.01$. (b) Bond orientational correlation functions. The fit is given by Eq.(\ref{Fit_g6_fluid}) and it is represented for $T=1.01$; inset in (a) shows the agreement between the fit and MC results for $g_6(s)$. (c) Bond orientational order parameter as function of the temperature, $<\phi_6>_{\mbox{\tiny MC}}$ is obtained with MC computations and  $<\phi_6>_g$ is the values obtained by fitting the tails of $g_6(s)$.  Inset : Histograms of $\phi_6$ for several temperatures.}
\label{Fig3}
\end{figure}
\begin{figure*}[ht]
\begin{center}
\centerline{\includegraphics[width=6.6in]{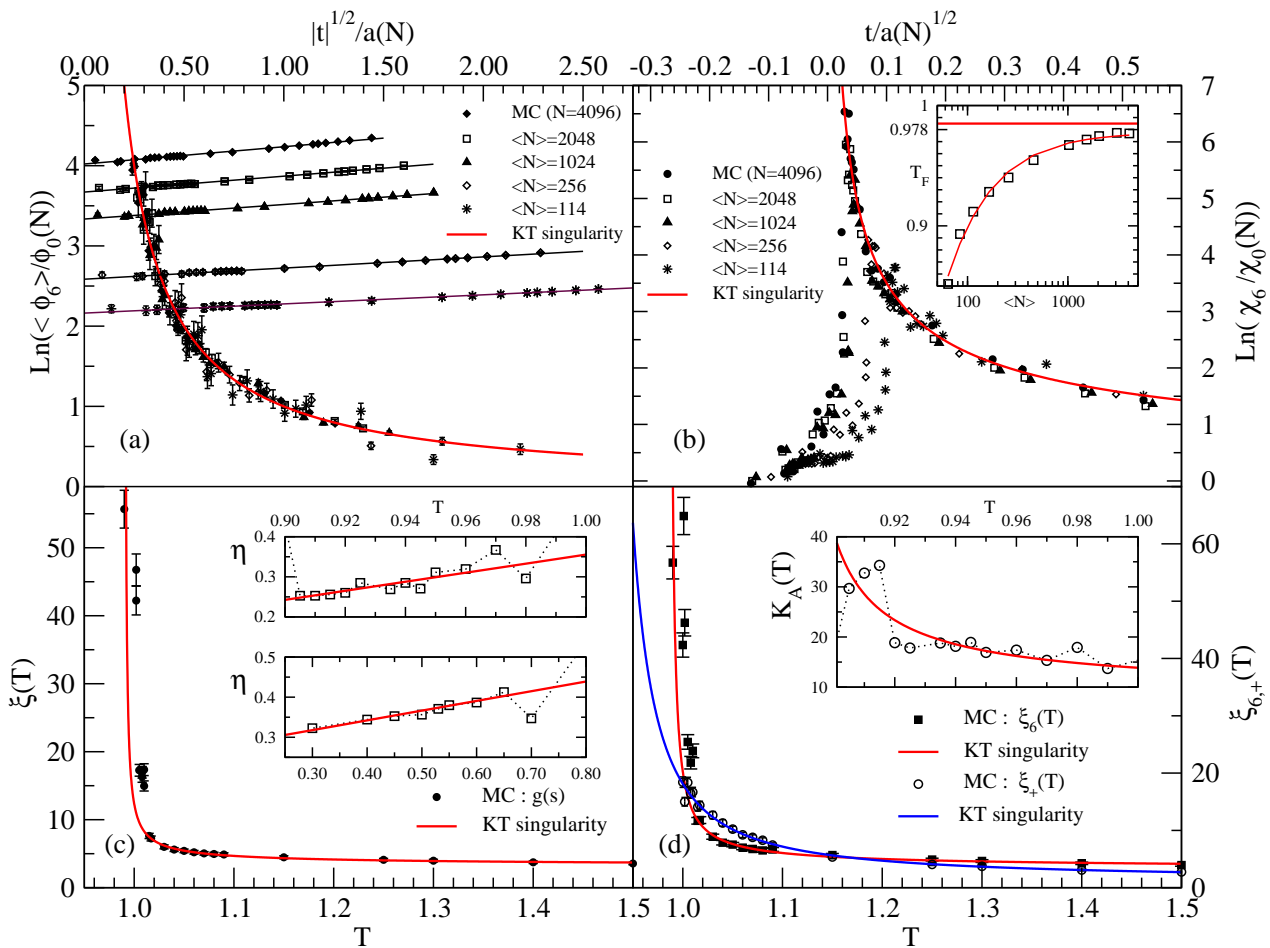}}
\end{center}
\caption{{\bf }(Color online) Dependences on temperature and Kosterlitz-Thouless singuralities for $n=0.71$ and $\rho=0.3$. (a) Finite size analysis of BOOP ; $t=(T-T_{F/H}(N))/T_{F/H}(N)$, the black straight lines correspond to $\ln<\phi_6> \simeq -0.412+0.37 \mid t\mid ^{1/2}$ for $t<0$. (b) Finite size analysis of the susceptibility. Inset : $T_{F/H}$ as function of the number of particles. (c) Correlation length $\xi(T)$ as function of temperature in the fluid phase. Inset : exponent $\eta(T)$ in the hexatic phase $0.9\leq T\leq1.0$ and in solid phase $T\leq 0.9$. (d) Bond orientational correlation lengths $\xi_6(T)$ and $\xi_+(T)$ as function of temperature in the fluid phase. Inset : stiffness $K_A(T)$ in the hexatic phase computed as $K_A(T)=18k_BT/\pi \eta_6(T)$.}
\label{Fig4}
\end{figure*}
\begin{figure}[ht]
\centerline{\includegraphics[width=3.3in]{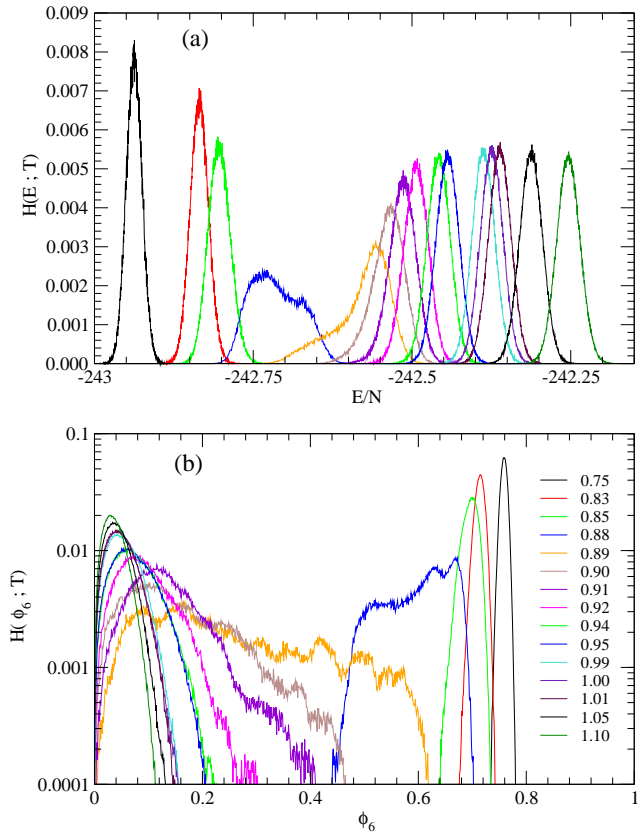}}
\vspace{-0.1in}
\caption{{\bf }(Color online) Histograms of energies and bond orientational order parameter $\phi_6$ for several temperatures with $n=1.65$ and $\rho=0.109$. The choice of $n$ corresponds to $n_H$ in Fig.\ref{Fig1} (a-inset). For each histogram shown, the value of the temperature is given in legend. The histograms are computed from the trajectories of the MC computations with $9\times 10^5$ and $2\times 10^6$ MC-cycles. These histograms of energies (trajectories) are used in the multiple histogram method to build the best estimate of the density of state. (a) Histograms of energy per particle. (b) Histograms of the order parameter $\phi_6$.}
\label{Fig5}
\end{figure}
\begin{figure*}[ht]
\begin{center}
\centerline{\includegraphics[width=6.6in]{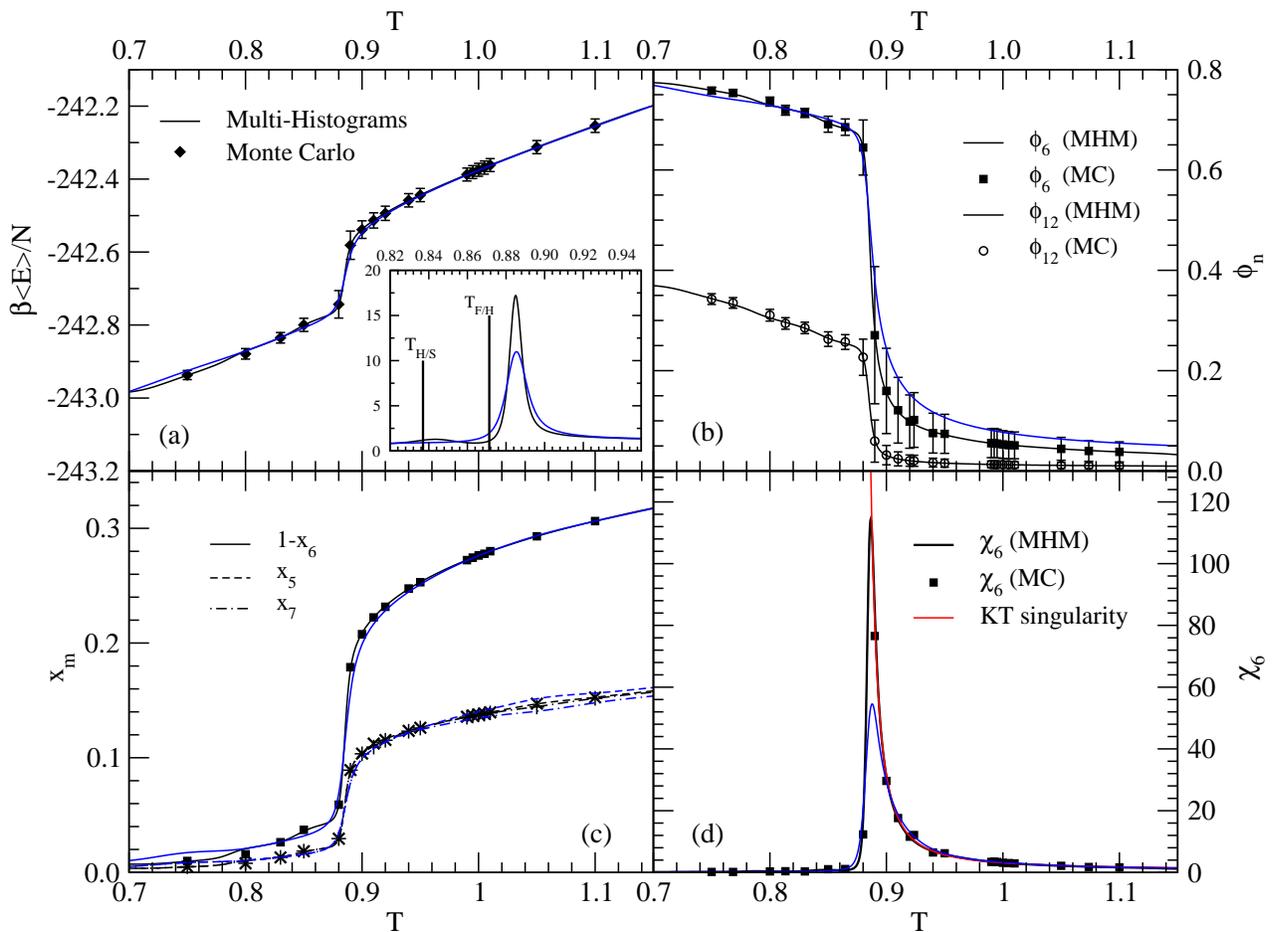}}
\end{center}
\caption{{\bf }(Color online) Interpolations of thermodynamic observables with the multiple histogram method as function of the temperature for  $n=1.65$ and $\rho=0.109$. All lines, but the KT singularity in (d), are computed with the multiple histogram method and symbol are averages computed with Monte Carlo simulations. Data in black are for $N=4096$ and curves in blue to $N=2025$ ; for clarity, the Monte Carlo averages for $N=2025$ are not represented on the figures. (a) Average energy per particle as function of $T$ (Inset : specific heat as function of T). (b) Bond orientational order parameter $\phi_6$ and $\phi_{12}$. (c) Interpolations of the fractions of $m$-coordinated particles computed with Voronoi constructions, $x_m$ are defined as $<N_m>/N$ where $N_m$ is the number of particles with exactly $m$ neighbors for a given configuration. The quantity $1-x_6$ is the fraction of defects. (d) Susceptibility of the order parameter $\phi_6$. The KT singularity is obtained by fitting the multiple histogram interpolation of the susceptibility by using Eq.(\ref{KT_sing}) with $\nu=1/2$, the fitted values are $X_0\simeq 0.322$, $a\simeq 0.823$ and $T_{F/H}\simeq 0.868$.}
\label{Fig6}
\end{figure*}
gives, with $\rho_{c}^{(F/H)}\simeq 0.891$ and $\rho_{c}^{(H/S)}\simeq 0.917$ \cite{Bernard:11}, $\Gamma_{\infty}^{(F/H)}\simeq 0.515$ and $\Gamma_{\infty}^{(H/S)}\simeq 0.529$. The transitions curves F/H and H/S obtained by using the empirical fits of Eq.(\ref{Fit_gamma}) are represented on FIG \ref{Fig1}(a). It is worthwhile to note that the transition curves as in Eq.(\ref{MeanField}) are equivalent to those found by Platzman and Fukuyama \cite{Platzman:74,Monarkha:book:03} for the Coulomb OCP monolayers when the Linedeman criterium is applied in the classical region.\\
On TABLE \ref{Table1}, we interpret the differences found for $n=3$ between MC computations and the experiments done by P. Keim and co-workers as follows. In the very nice experiments done in refs.\cite{Keim:07}, the superparamagnetic colloids are confined at an air/water interface and dipoles are induced by an external magnetic field ; therefore a small fluctuation in the vertical position of the particles would result in a small attractive contribution in the dipolar interactions, this would make the ordered phases (Hexatic and Solid) stable at smaller densities, smaller values of the coupling constant. This may explain the small quantitative differences, but an excellent qualitative agreement, found between superparamagnetic colloids confined at an air/water interface and pure numerical monolayer systems with $1/r^3$ interactions. Moreover, in the Molecular Dynamics done by S. Lin and co-workers \cite{Lin:06}, the coupling constants for both transitions, Fluid/Hexatic and Hexatic/Solid, that they found in their analysis agree very well with the ones found in the present work done with Monte Carlo computations and Ewald sums. Therefore, it seems reasonable to consider that the real interaction between the superparamagnetic colloids in an external magnetic field deviate slightly form a true $1/r^3$ potential ; one should note also that the experiments and the simulations are in agreement with the small amplitude $\Delta\Gamma \sim 3$ in which the Hexatic phase can be observed for $n=3$.\\
With the empirical curves given by Eqs.(\ref{Fit_gamma},\ref{MeanField}) and the numerical value of the $\Gamma_i$-parameters, we found $\rho_c^{(F/H)} > \rho_c^{(H/S)}$ for $4.1\leq n\leq 127$ with $T=1.0$ and $Q=14$. This indicates that the hexatic phase would disappear in this range of $n$ and that the KTHNY transitions are replaced by a first order phase transition \cite{Chui:82}.\\
On FIG.\ref{Fig2}, we show snapshots of monolayer systems in Hexatic and Hexagonal phases near melting (Runs B). For $n=0.71$ (FIG.\ref{Fig2}(a)), the disclinations are bound in pairs (equivalent to isolated dislocations) and quartets and they are more or less uniformly distributed in the monolayer, this is consistent with the KTHNY theory \cite{Nelson:book:83,Nelson:79} and the system is in the fluid phase very near to the hexatic transition for the thermodynamic parameters given in the caption of FIG.\ref{Fig2}(a). For $n=7.30$ (FIG.\ref{Fig2}(b)), the defects form small lines and aggregate ; these arrangements of the defects may be considered as seeds for grain boundaries in a first order melting \cite{Chui:82,Nosenko:09}. Qualitatively, the range of the interactions can help us to understand the differences found in the distribution of defects. In one hand, for long ranged interactions ($n=0.71$), each particle interacts with all particles, thus it is energetically unfavourable for the defects to aggregate in lines. On the other hand, for much shorter interactions ($n=7.30$), heterogeneous distributions of defects are not averaged out by the range of interactions ; on contrary, the quite short range of the interaction favours the growth of grain boundaries.
\begin{figure}[ht]
\centerline{\includegraphics[width=3.3in]{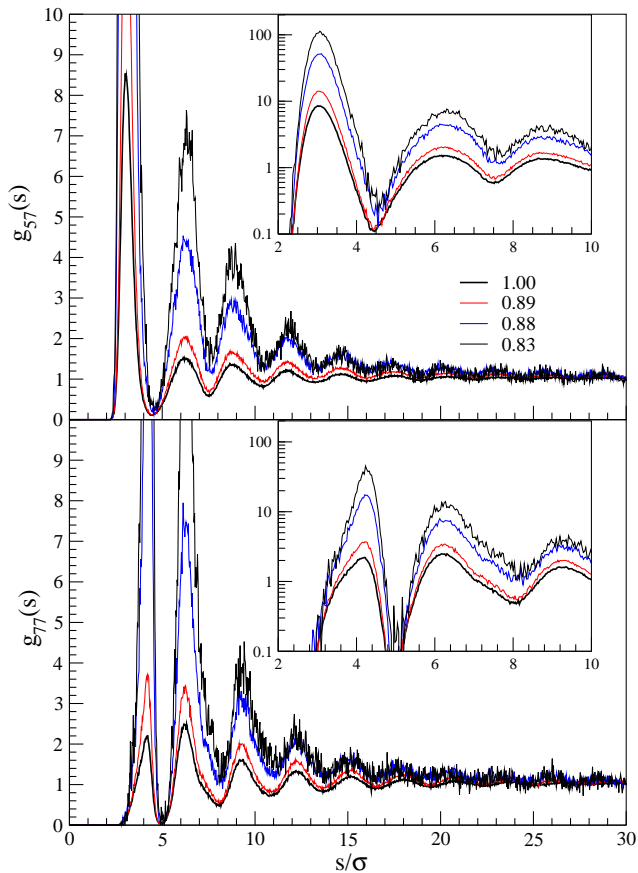}}
\vspace{-0.1in}
\caption{{\bf }(Color online) Center to center correlation functions $g_{pq}(s)$ for pair of particles with $p$ neighbors and $q$ neighbors for  $n=1.65$ and $\rho=0.109$. For the $g_{57}(s)$ correlation function (top), at $T=0.83$, the first two peaks are approximately located at $s=3.1\mbox{ }\sigma$ (height : $\sim 110$) and $6.1\mbox{ }\sigma \leq s \leq 6.6\mbox{ }\sigma$ (height : $\sim 7$). The first two peaks in the $g_{77}(s)$ correlation function (bottom) are located at $s=4.2\mbox{ }\sigma$ (height : $\sim 45$) and $6.0\mbox{ }\sigma \leq s \leq 6.6\mbox{ }\sigma$ (height : $\sim 12$).}
\label{Fig7}
\end{figure}
\begin{figure*}
\begin{tabular}{ccc}
(a)\includegraphics[width=2.in]{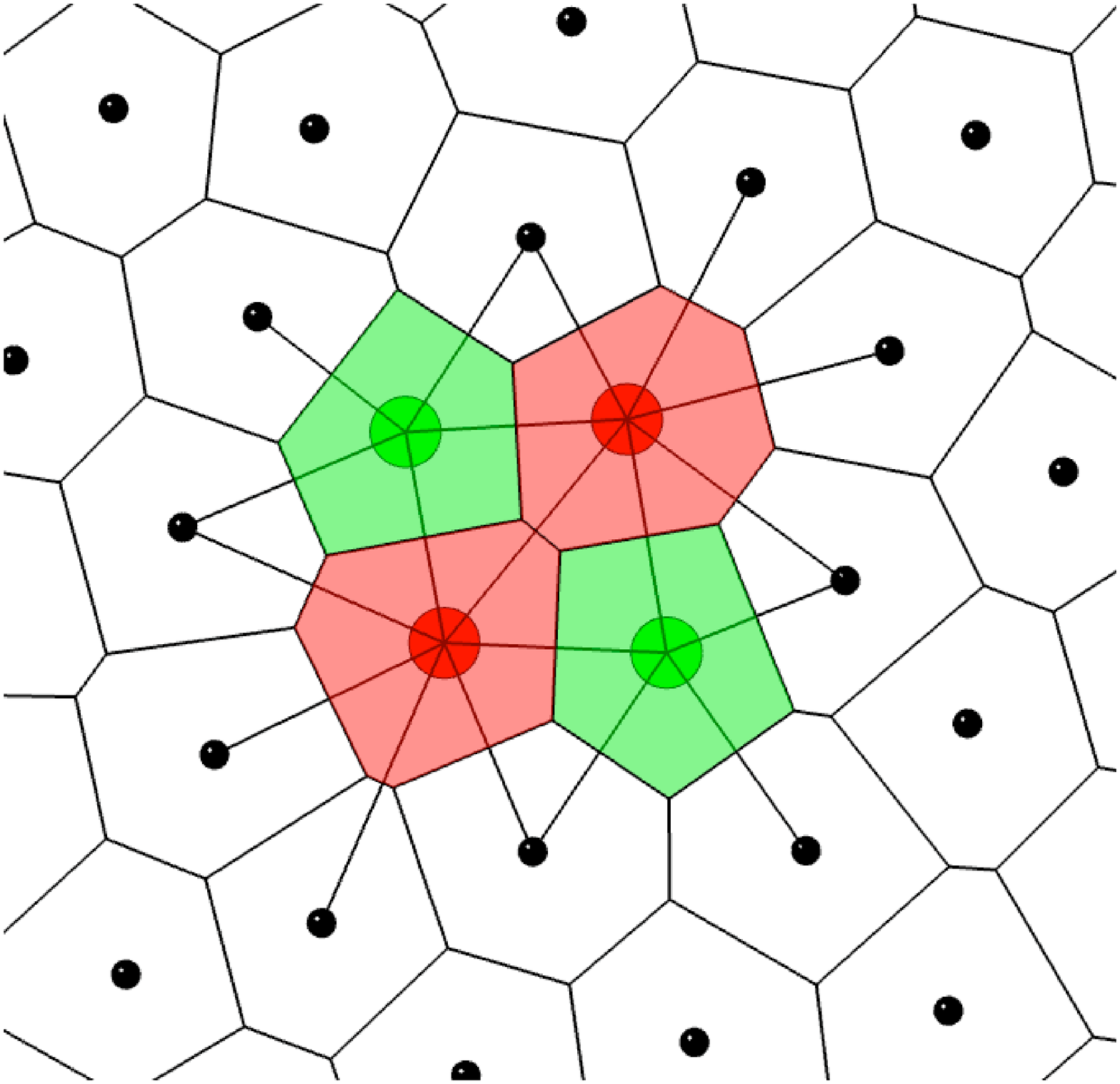} & (b)\includegraphics[width=2.in]{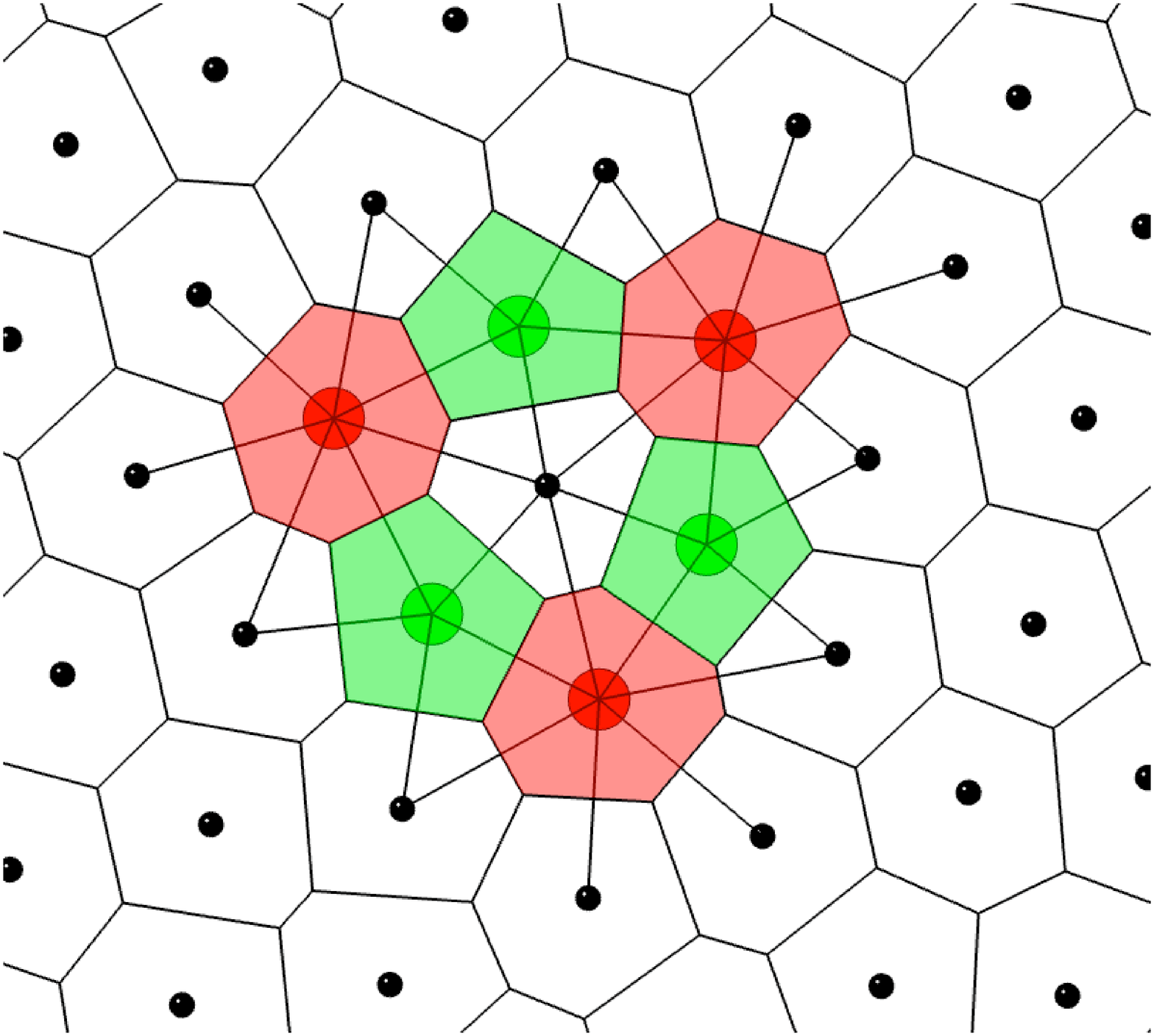} & (c)\includegraphics[width=2.3in]{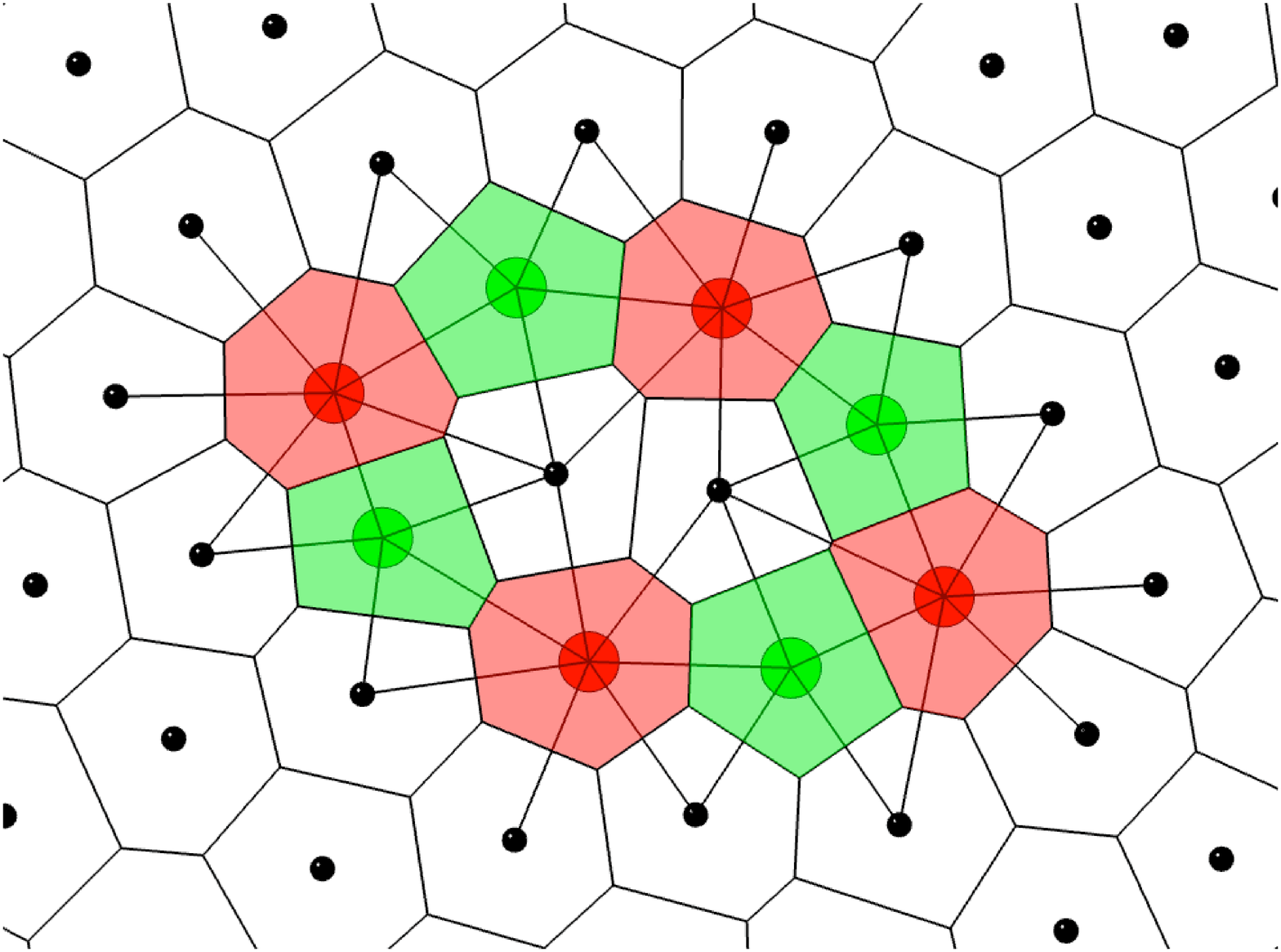}\\
&\\
(d)\includegraphics[width=2.in]{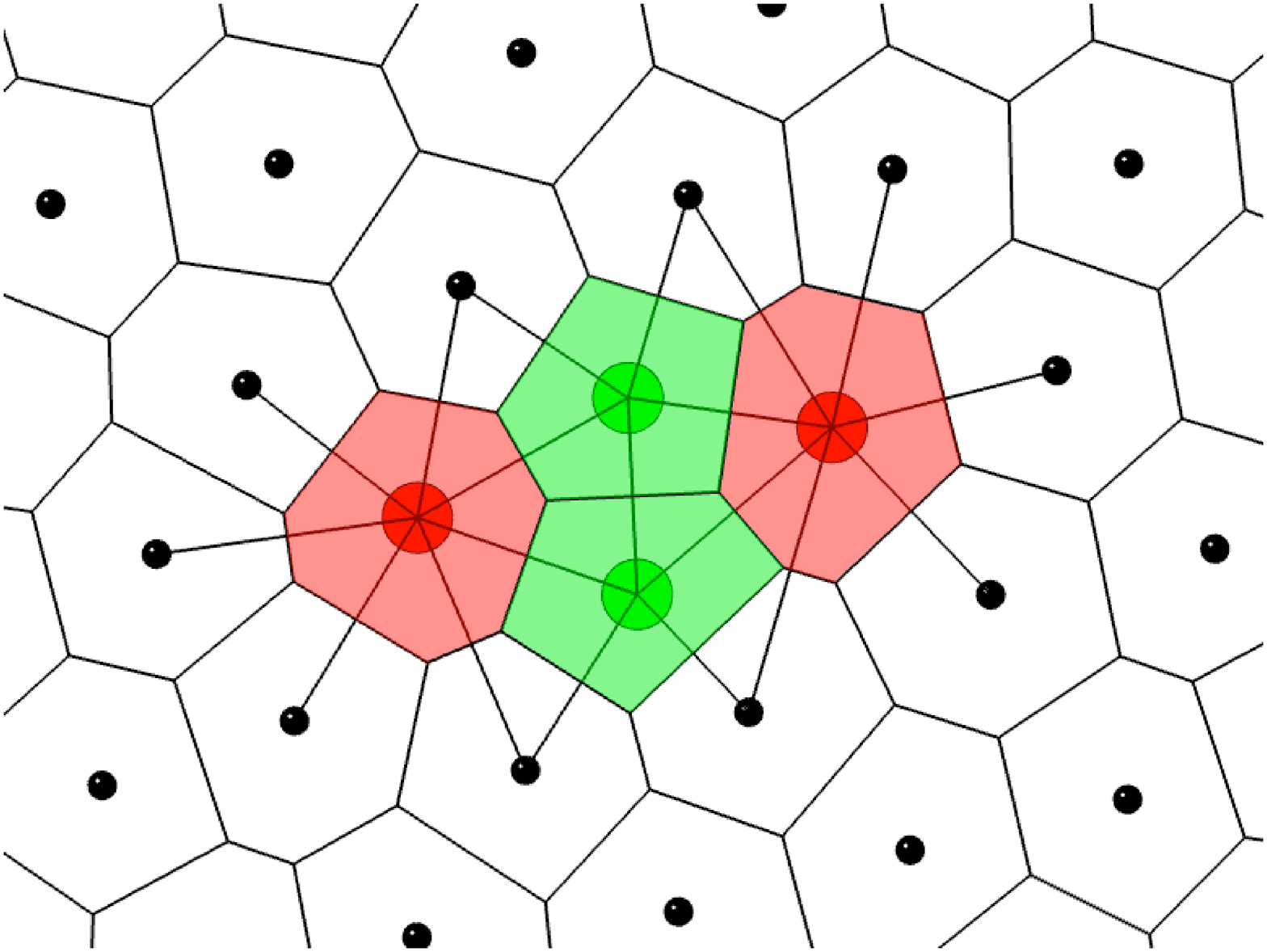} & (e)\includegraphics[width=2.in]{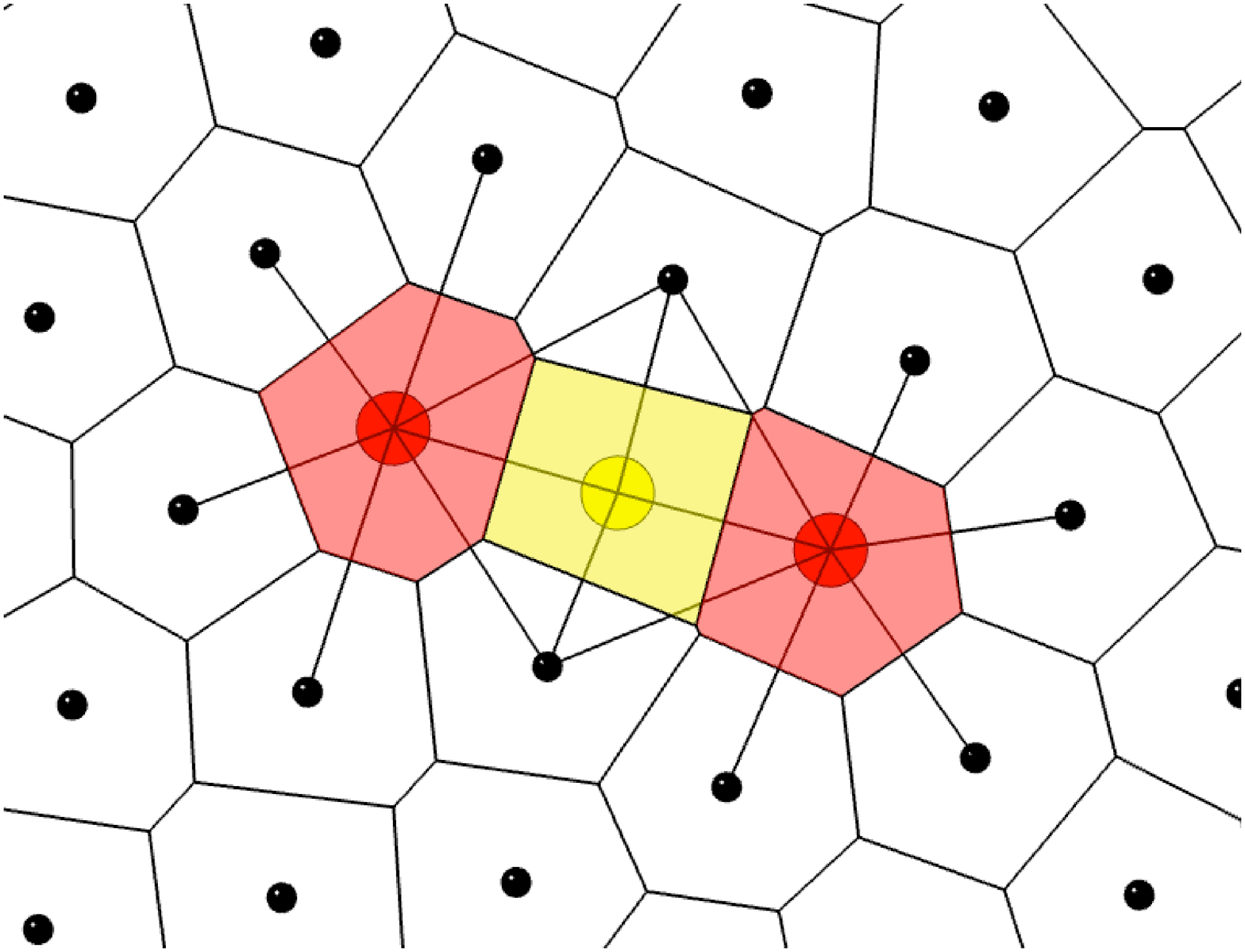} & (f)\includegraphics[width=2.in]{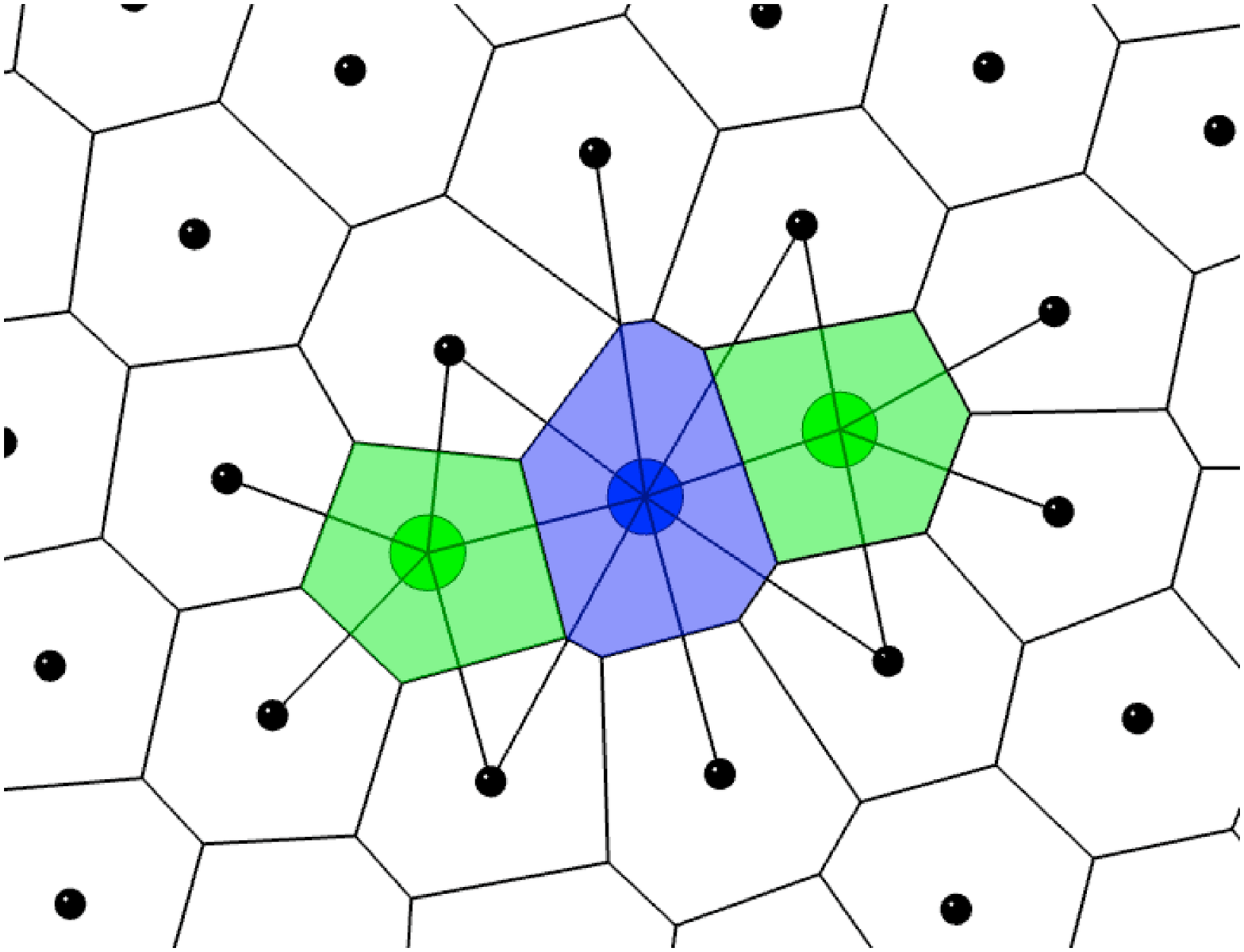}
\end{tabular}
\caption{{\bf }(Color online) Topological defects in the Hexatic (near Solid) phase ; $n=1.65$, $T=0.83$ and $\rho=0.109$. (a,b,c) Quite frequent defects ; (d,e,f) rare defects. In all snapshots, the diameter of the coloured disks is $\sigma (\equiv 1)$. The Delaunay construction is represented only for particles that do not have 6 neighbors. The Voronoi cells are represented for all particles. The color code for the particles is defined by the number of neighbors : 4 neighbors (yellow), 5 (green), 7 (red), 8 (blue). The small black disks inside white Voronoi cells are the point particles with 6 neighbors.}
\label{Fig8} 
\end{figure*}
\begin{figure*}
\begin{tabular}{cc}
(a)\includegraphics[width=3.6in]{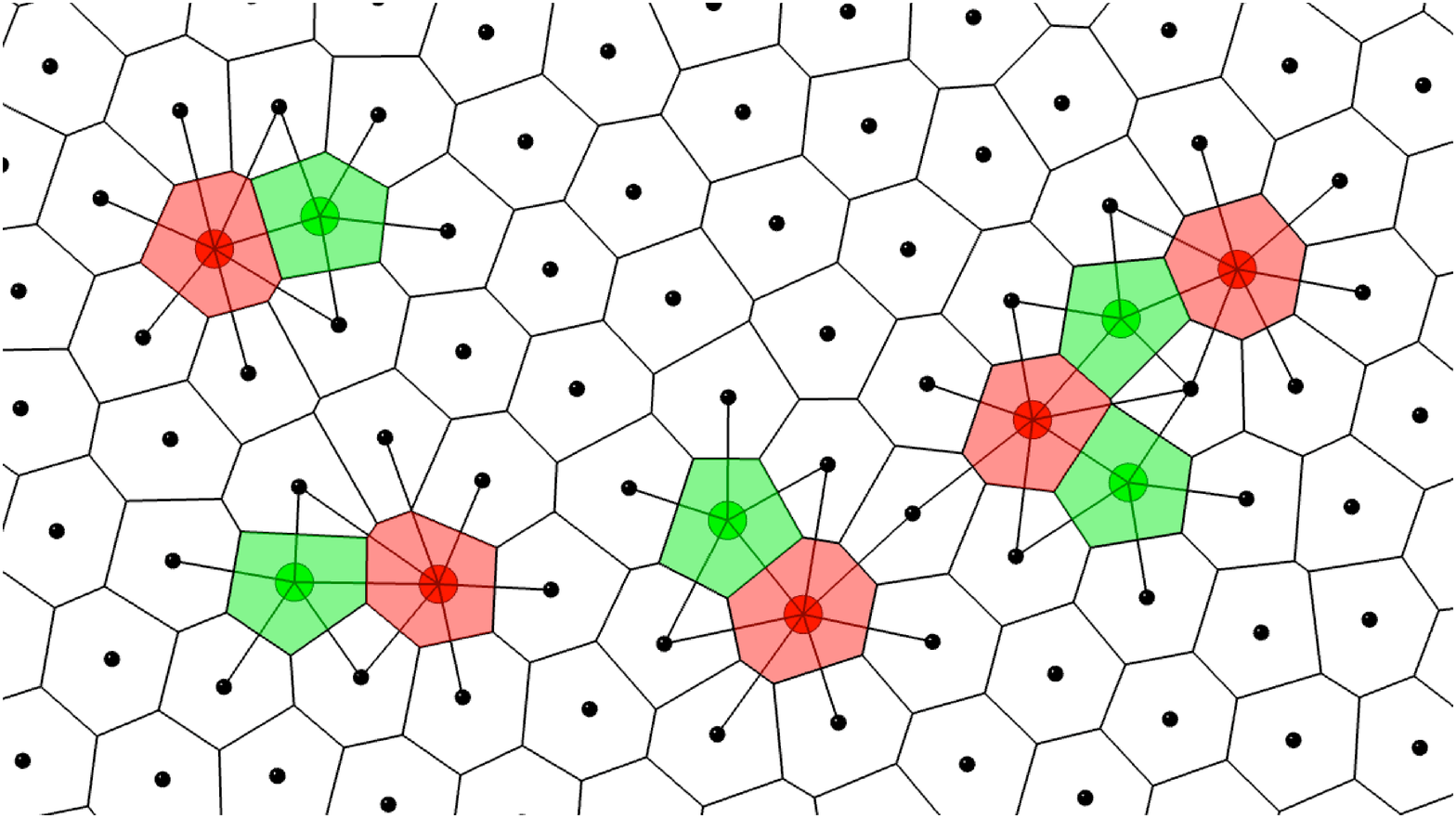} & (b)\includegraphics[width=2.6in]{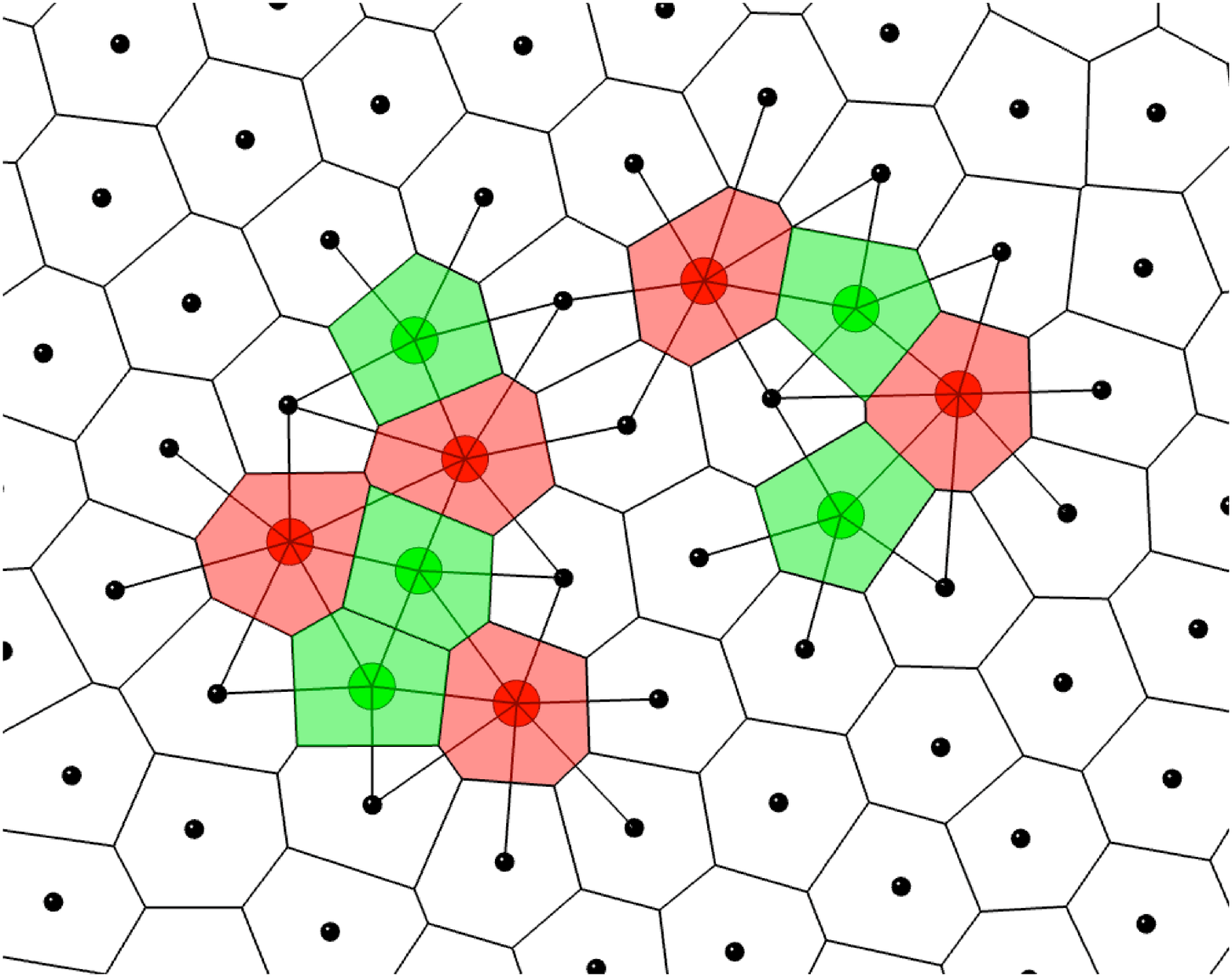}
\end{tabular}
\caption{{\bf }(Color online) Topological defects in the Hexatic (near Fluid) phase ; $n=1.65$, $T=0.88$ and $\rho=0.109$. }
\label{Fig9} 
\end{figure*}
\section{Monte Carlo computations and analysis.}
In FIG \ref{Fig1}.(b), we represent BOOP for Runs A and B, and on FIG.\ref{Fig3}(c) they are represented as function of temperature for $n=0.71$ and $\rho=0.3$.\\ 
The insets (c,d) in FIG \ref{Fig1}.(b) show straight line fits of the fourth order cumulants in the region of transition for several size of subsystems. For Runs A, FIG \ref{Fig1}.(c), the straight line does not cross in a single point ; this argues in favour of a first order transition for the melting of the hexagonal 2D solid phase \cite{Gribova:11}. For Runs B, FIG \ref{Fig1}.(d), the shift in the crossing of the straight line fits of the cumulants is less apparent ; in consideration with the other results described in this paper (and done by others in previous studies), it seems plausible that the transition is weakly first order or located quite close to a kind of multicritical point separating the KTHNY theory from a first order phase transition \cite{Zippelius:80}.\\   
On Figs.\ref{Fig3}(a,b), the red curves are fits obtained with Eqs.(\ref{Fit_g6_fluid},\ref{Fit_g6_hexatic}). The values of $<\phi_6>_g$, as well as the Monte-Carlo averages $<\phi_6>_{\mbox{\tiny MC}}$ are reported on Fig.\ref{Fig3}(c). Both evaluations of the bond orientational order parameter agree extremely well in hexatic and solid hexagonal phases, but a significant difference is seen in the fluid phases. This difference in the fluid phase is well understood since bond orientational order in the fluid phase is short ranged and $<\phi_6>_{\mbox{\tiny MC}}$ is the average over all local order parameters, while $<\phi_6>_g$ is the asymptotic value of $g_6(s)$. In the solid phase $T<0.90$, the center to center correlation functions $g(s)$ are exactly those of a hexagonal lattice with thermal fluctuations and for $s>10\mbox{ } \sigma$, the bond orientational correlation functions $g_6(s) \simeq <\phi_6>_g =<\phi_6>_{\mbox{\tiny MC}}$ within the statistical fluctuations, this indicates a long range of the bond orientational order. The dependence on temperature for the histograms of $\phi_6$ (Inset in FIG.\ref{Fig3}(c) - $n=0.71$) is similar to the dependence found experimentally in superparamagnetic colloids ($n=3$) confined at an air/water interface (cf. FIG.6 of ref.\cite{Keim:07}). For $T=1.00$ and $T=1.01$, the histograms of $\phi_6$ are very broad, it corresponds to the KT-singularity of $\chi_6$ ; similar behaviors are observed on histograms computed for $n=1.65$ given in FIG.\ref{Fig5}(b).\\
On FIG.\ref{Fig4}, we summarize the analysis on the done on Kosterlitz-Thouless essential singularities to obtain the transition temperatures $T_{F/H}$ and $T_{H/S}$ for a set of Monte Carlo computations ($n=0.71$ and $\rho=0.3$). All singularities showed on FIG.\ref{Fig4} are plotted with $\nu=1/2$ ; however, the numerical results given in FIG.\ref{Fig4} may also be quite well represented with $0.35\leq \nu\leq 0.55$. In KTHNY theory, the essential singularities as in Eq.(\ref{KT_sing}) occur with $\nu=1/2$ (F/H) or with $\bar{\nu}=0.369\mbox{ }634 ...$ (H/S) \cite{Young:79,Nelson:book:83,Kosterlitz:74,Nelson:79,Strandburg:88} (or with $\nu=2/5$ \cite{Nelson:78}). The values of the transition temperatures obtained by using $\nu=1/2$ or $\bar{\nu}$ are in the uncertainties ; for instance, for $\chi_6$ in FIG.\ref{Fig4}(b) with $\nu=1/2$ one finds $T_{F/H}=0.9783$ and with $\bar{\nu}$ one has $T_{F/H}=0.9845$, and both fits are equally good.\\
At the Fluid/Hexatic transition, the KTHNY theory states that $\nu=1/2$, this is fully consistent with the Monte Carlo results obtained in the present work ; however, we are not able yet to estimate with a convenient accuracy the value of the exponent $\nu$ in Eq.(\ref{KT_sing}) at the Hexatic/Solid transition and verify, as stated in KTHNY theory, that $\nu(H/S)=\bar{\nu}\simeq0.369\mbox{ }634 ...  $.\\ 
On FIG.\ref{Fig4}(a,b), we give finite size analysis for BOOP and susceptibility represented as function of the scaling variable $t=(T-T_{F/H}(N))/T_{F/H}(N)$ and, on FIG.\ref{Fig4}(c,d), we represent as function of the temperature $\xi(T)$, $\xi_+(T)$, $\xi_6(T)$, $\eta(T)$ and $\eta_6(T)$ obtained by fitting the tail of the correlation functions. The numerical values of $T_{F/H}(N)$ used in the definition of the scaling variable $t$ are obtained from Eq.(\ref{KT_sing}) for each subsystem size.\\
The collapse of data for different subsystem sizes on the universal Kosterlitz-Thouless essential singularity for both the bond orientational order parameters and susceptibilities, FIG.\ref{Fig4}(a,b), and also the dependence on temperature of correlation lengths, FIG.\ref{Fig4}(c,d), are strong arguments in favour of the KTHNY theory for $n=0.71$ and $\rho=0.3$. Same results are obtained for Runs A and B (cf. TABLE \ref{Table1}) for $n<2$ and also for computations with $n=1.48$, $1.65$ and $3$.\\
For $n=0.71$, $\rho=0.3$ and $\tilde{\Gamma}=196$, compiling the results from finite-size analysis and KT-singularities, we found that the Fluid/Hexatic transition is located at temperature $T_{F/H}=0.986\pm 0.008$. From the KT-singularities found for the stiffness $K_A(T)$ (inset FIG.\ref{Fig4}(d)) and $\xi_+(T)$ (blue line on FIG.\ref{Fig4}(d)), we found the temperature of the Hexatic/Solid transition at $T_{H/S}=0.913\pm 0.011$. From the singularity found for $K_A(T)$, we compute $\eta_6(T_{H/S})=0.385$, that is different from the value $1/4$ predicted in the KTHNY theory ; we may interpret this disagreement as it is done for Coulomb interaction \cite{Clark:09}, by arguing that the critical exponents depend on the range of the interaction \cite{Fisher:72,Luijten:02}. More precisely, we consider that the periodic boundary conditions, which introduce an arbitrary long range positional order between particles, and, more fundamentally, the dependence of critical exponents with the range of the interaction potentials \cite{Fisher:72,Luijten:02,Clark:09} may explain the difference between the KTHNY predictions in the Hexatic phase and the observed critical behavior of $\xi(T)$ and values of $\eta(T)$ and $\eta_6(T)$ reported on insets of FIG.\ref{Fig4}(c,d). A complete study of the dependence of the critical exponents with $n$ is needed to understand precisely the critical behavior of the $n$-OCP monolayer systems and modifications induced by the range of interactions in the KTHNY predictions ; such a study, would be of great interest in the understanding of the melting in two dimensions.\\ 
As shown in the following, the Kosterlitz-Thouless essential singularity can be very well represented with the help of the multiple histogram method (MHM) \cite{Ferrenberg:88,Newman:book:99}. On FIG.\ref{Fig5}, we represent the energy and the order parameter histograms for the temperatures used in the multiple histogram reweighting method the system with $N=4096$ particles for $n=1.65$ and $\rho=0.109$. The histograms shown on FIG.\ref{Fig5} are computed from the trajectories of the Monte Carlo simulations. The overall qualitative features of the order parameter histograms for both $n=1.65$ (FIG.\ref{Fig5}(b)) and $n=0.71$ (Inset of FIG.\ref{Fig3}(c)) are similar and in agreement with the ones found for superparamagnetic colloids confined at an air/water interface \cite{Keim:07} (see also FIG.2 of ref.\cite{Gribova:11}). It is also worthwhile to note that, close to the KT-singularity ($T=0.90-0.88$) the order parameter histograms broaden strongly and tend to become quite flat between for $\phi_6$ between $0.1$ and $0.6$, making the susceptibility $\chi_6$ violently increasing. From the MC trajectories used to compute the histograms of FIG.\ref{Fig5}, we apply the multiple histogram reweighting technique to obtain interpolations of the thermodynamic observables at temperatures other than the ones used in the Monte Carlo computations. These interpolations are shown on FIG.\ref{Fig6} for systems with 4096 particles (black curves and symbols) and 2025 particles in blue.\\
On FIG.\ref{Fig6}(a), we represent the average energy per particle as function of $T$ ;  the variation of the energy above $T_{F/H}$ is due to the gradual dissociation of disclination pairs (see also FIG.\ref{Fig6}(c)). In the inset we represent the specific heat computed with the Multiple Histogram Method for both system size $N=4096$ and $N=2025$, the transition temperatures $T_{F/H}$ and $T_{H/S}$ found in KT-singularities are represented as vertical lines. The peak in the specific heat near the Fluid/Hexatic transition occurs above $T_{F/H}$ and it is also due to gradual dissociation of disclination pairs in agreement with the KTNHY theory \cite{Nelson:book:83,Strandburg:88} ; this peak is similar to the one found because of the vortex contribution in the $XY$-model according to which the height, position and shape of the maximun are not universal \cite{Nelson:book:83,Tobochnik:82}. On the system with 4096 particles, a small bump is seen also on the specific heat just before $T_{H/S}$, it would correspond to the dissociation of dislocation pairs.\\
On FIG.\ref{Fig6}(b), we represent the bond orientational order parameters $\phi_6$ ($N=4096$ and 2025) and $\phi_{12}$ ($N=4096$) and on FIG.\ref{Fig6}(d), the susceptibility $\chi_6$ is shown for both system sizes. The effects of the finite size of the system (and periodic boundary conditions) can be seen on both figures and these finite size effects are consistent with the theory \cite{Privman:book:90}. The maximum found for the susceptibility with the multiple histogram method is very slightly shifted to the higher temperature with decreasing the system size. More precisely, for the system with $N=4096$ particles the maximum is located at $T=0.886$ and $\chi_6(T=0.886\mbox{ } ; 4096)=114.3$ while for the system with $N=2025$ particles the maximum is at $T=0.888$ and $\chi_6(T=0.888\mbox{ }; 2025)=54.6$, this agrees well with the finite size analysis done on subsystems (cf. FIG.\ref{Fig4}(b) - symbols for $N=4096$ and $<N>=2048$).\\
On Fig.\ref{Fig6}(d), the red curve is the KT-singularity, Eq.(\ref{KT_sing}), obtained by fitting the multiple histogram interpolation of the susceptibility, the agreement is very good both with the interpolation done with the multiple histogram reweighting and with the finite size analysis on subsystems.\\
On FIG.\ref{Fig6}(c), the fraction of $m$-coordinated particles with $m=5$ and $7$ and the fraction of defects, defined as $1-x_6$, are shown. On all the temperature range used in the multiple histogram interpolation, one has $x_5= x_7$ with an excellent accuracy in the solid and hexatic phase and we found $1-x_6=x_5+x_7$. In the fluid, there are few particles with $m=4, 8$ and $9$ neighbors, however, the equality  $x_5= x_7$ still holds to about 1\% accuracy. The fractions of $m$-coordinated particles with $m=4,8$ and $9$ are in the ordered phases (hexatic and solid) : $x_4 < 4\times 10^{-5}$ ; $x_8< 2\times 10^{-5}$ and $x_9< 10^{-9}$ ; and, in the fluid phase : $x_4\simeq 7\times 10^{-4}$, $x_8\simeq 10^{-3}$ and $x_9< 10^{-6}$. From the fraction of defects in the Solid phase, we may estimate the core energy $E_c$ at low temperature by using 
\begin{equation}
\label{fraction_d}
\displaystyle 1-x_6=X_d^0\exp\left(-2 E_c(n)\frac{T_{H/S}}{T}\right)
\end{equation}
to model the defect fraction as an Arrhenius law in the low temperature limit \cite{Tobochnik:82,Murray:87,Sengupta:00,Wierschem:11}.\\
With Eq.(\ref{fraction_d}), we found : for $n=1.48$, $E_c=4.38$ ; $n=1.65$, $E_c=5.02$ and for $n=3.00$, $E_c=3.46$. The change from the KTHNY melting to the grain boundary induced melting is predicted to occur when $E_c < 2.84$ \cite{Chui:82,Saito:83,Strandburg:88,Sengupta:00} ; therefore, the results found for $E_c$ from the analysis of the fraction of defects support the KTHNY melting according to the Chui-Saito criterium \cite{Chui:82,Saito:83} (see also next section).\\ 
The center-to-center correlation functions $g_{57}(s)$ and $g_{77}(s)$ are represented on FIG.\ref{Fig7} for the Fluid and Hexatic phases and near the Hexatic/Solid transition ($T=0.83$). As explained in Sect.II, in Solid and  Hexatic phases the fraction of defects is small, thus large statistical fluctuations are found in $g_{57}(s)$ and $g_{77}(s)$ for $s> 10\sigma$ ; for instance, at $T=0.83$, one has $x_5=x_7=(1.3\pm 0.3)\times 10^{-2}$.\\
The first peak in the $g_{57}(s)$ gives an estimate of the strength of the bound between the 5 and 7-coordinated particles (positive and negative disclinations) ; at $T=0.88$ (below $T_{F/H}$ - cf. the KT-singularity of $\chi_6$), the first peak is located at $s = 3.1$ $\sigma$ and the height is about $52$ and, just above $T_{F/H}$, at $T=0.89$, the height of the first peak falls to 14.2. In the Fluid phase ($T=1.00$), the first peak is located at $s = 3.0$ $\sigma$ and its height is about 8.5.\\
Similarly, the height of the first peak in the  $g_{77}(s)$ gives an estimate of the strength of the bound between dislocations (see FIG.\ref{Fig8} (a) and (d)). At $T=0.83$, in the Hexatic phase very close to the Hexatic-Solid transition, the first peak is located at $s = 4.2$ $\sigma$ and its height is about 45.1, it falls to 17.5, in the Hexatic phase close to the Fluid-Hexatic transition ($T=0.88$) and to 3.7 in the Fluid phase at $T=0.89$.\\ 
For both correlation function $g_{57}(s)$ and $g_{77}(s)$ (and also $g_{55}(s)$ - not shown), the locations of the first peak are extremely well represented with the topological defect shown on FIG.\ref{Fig8}(a), that corresponds to a pair of dislocations or a quartet of disclinations. Thus, the analysis of the behaviour of the first peaks of the correlations functions $g_{57}(s)$ and $g_{77}(s)$ with an increase of the temperature supports the view of the KTHNY theory that describes the melting from Solid to Hexatic as a dislocation-unbinding transition, followed by the Hexatic to Fluid transition stemming from the disclination-unbinding transition \cite{Strandburg:88}.\\
More generally, the first three peaks of $g_{57}(s)$ and $g_{77}(s)$ are very well located by taking into account only the defects of FIG.\ref{Fig8} (a-c). The defects shown in FIG.\ref{Fig8} (b) and (c) are quite frequent in the Hexatic phase, however, as it can be seen on the snapshot given in FIG.\ref{Fig2}(a), other topological arrangements of disclinations and dislocations with irregular shapes can also be found, some are given in FIG.\ref{Fig9}. The defects shown on FIG.\ref{Fig8} (d-f) are very rare.    
\section{Discussion and Perspectives}
In the KTHNY theory, the melting of the hexagonal solid in two dimensions is described by two continuous transitions. From the hexagonal Solid phase to the isotropic Liquid phase, the first transition is induced by the unbinding of the dislocations and it corresponds to the Solid-Hexatic transition, the second transition is the disclination-unbinding transition from the critical Hexatic phase to the isotropic liquid \cite{Nelson:78,Nelson:book:83,Strandburg:88}.\\ 
From the continuum elasticity theory, the dislocation-unbinding transition is described with the Hamiltonian \cite{Nelson:78} 
\begin{equation}
\label{H_disloc}
\begin{array}{ll}
\displaystyle H_{\mbox{\tiny disloc}} &\displaystyle =\frac{K}{8\pi}\sum_{\bm{r}\neq\bm{r}^{\prime}}\left[\frac{\bm{b}(\bm{r})\cdot(\bm{r}-\bm{r}^{\prime})\bm{b}(\bm{r}^{\prime})\cdot(\bm{r} - \bm{r}^{\prime})}{\mid \bm{r}-\bm{r}^{\prime} \mid^2}\right. \\
&\\
&\displaystyle - \left.\bm{b}(\bm{r})\cdot\bm{b}(\bm{r}^{\prime})\mbox{ }\ln\left(\frac{\mid \bm{r}-\bm{r}^{\prime} \mid}{a_c}\right)  \right ]\\
&\\
&\displaystyle +E_{c}\sum_{\bm{r}}\mid \bm{b}(\bm{r})\mid^2
\end{array}
\end{equation}
with $\bm{b}(\bm{r})$ the Burgers vector of a dislocation located at $\bm{r}$, $K$ the Young's modulus, $a_c$ the core radius that defines the distance outside of which the dislocation energy obey the logarithmic form of the previous equation and $E_c$ is the core energy that gives the contribution to energy inside the core radius.\\
The core energy $E_c$ has a crucial influence on the mechanisms of melting in two dimensions \cite{Chui:82,Saito:82,Saito:83}. Renormalization group analysis of the KTHNY theory uses the fugacity of defects, $y=\exp(-E_c)$, in the recursive relations \cite{Nelson:book:83,Sengupta:00}. In the analytical description of the grain-boundary theory of melting \cite{Chui:82}, S.T. Chui predicts that the melting is weakly first order when $E_c < 2.84$ and becomes strongly first order as $E_c > 2.84$. However,   Monte-Carlo simulations using the Hamiltonian of Eq.(\ref{H_disloc}) done by Y. Saito \cite{Saito:82,Saito:83} showed that the dislocation-unbinding transition is well described by the KTHNY theory for large $E_c$ while it became first order for small value of $E_c$, because of the nucleation of grain boundary loops. In the grain-boundary induced melting theory, it is assumed that the distance between dislocations making a grain boundary is small between grain boundary, this approximation breaks down for large value of $E_c$ for which the distance between dislocation pairs become large \cite{Strandburg:88} (see also FIG.\ref{Fig2} (a) and (b)). Therefore, the core energy $E_c$ can be used to define a criterium for the dislocation-unbinding transition : for $E_c < 2.84$, the transition is (weakly) first order and for $E_c > 2.84$ it follows KTHNY theory.\\
In experiments and in computer simulations, we may obtain an estimate of $E_c$ from the fraction of defects in the solid phase \cite{Murray:87,Tobochnik:82,Sengupta:00,Wierschem:11} by using an Arrhenius law as in Eq.(\ref{fraction_d}). The estimates of $E_c$ done in the previous section for the $n$-OCP monolayer show first that, for $n\leq 3$, the dislocation-unbinding transition follows KTHNY theory according to the Chui-Saito criterium, and, second, that $E_c$ depends on $n$. In the present work, the estimates of $E_c$ are given for only three values of the power $n$ ; this does not permit yet to obtain the dependence of  $E_c$ on $n$ and to improve the Chui-Saito criterium.\\
A dislocation is equivalent to a pair of disclinations strongly bounded. As shown by B.I. Halperin and D.R. Nelson \cite{Nelson:79}, the disclination-unbinding transition can be described by using an isomorphism with the $XY$ model \cite{Kosterlitz:74} and the Coulomb gas in two dimensions \cite{Young:79}, with the disclinations represented as vortices or charges. The Hamiltonian for the disclination-unbinding transition is \cite{Nelson:79,Nelson:book:83,Strandburg:88} 
\begin{equation}
\label{H_disc}
\begin{array}{ll}
\displaystyle H_{\mbox{\tiny disc}} &\displaystyle =-\frac{\pi K_{A}}{36}\sum_{\bm{r}\neq\bm{r}^{\prime}} s(\bm{r})s(\bm{r}^{\prime})\mbox{ }\ln\left(\frac{\mid \bm{r}-\bm{r}^{\prime} \mid}{a_{cd}}\right)\\
&\\
&\displaystyle +E_{cd}\sum_{\bm{r}} s(\bm{r})^2
\end{array}
\end{equation}
with $K_A$ the Frank constant (the coupling constant related to the distortion of the bond-angle field), $s(\bm{r})$ the charge of the disclination located at $\bm{r}$, $a_{cd}$ is the disclination core size and $E_{cd}$ the disclination core energy. In the isomorphism with the Coulomb gas, the charges assigned to disclinations can be chosen as $s(\bm{r})=+1$ for a particle with seven nearest neighbors (Voronoi cells and particles in red in FIGS.\ref{Fig2}, \ref{Fig8} and \ref{Fig9}) and $-1$ for five nearest neighbors (represented in green). The analysis of the first peaks of the correlation functions $g_{57}(s)$ and $g_{77}(s)$ (cf. FIG.\ref{Fig7}) done in the previous section is in excellent qualitative agreement with the disclination-unbinding transition for the Hexatic and Fluid phases.\\
The results of the Kosterlitz-Thouless theory for the $XY$ model and Coulomb gas in two dimensions can be used for the disclination-unbinding transition, the Hexatic/Fluid transition. In the $XY$ model, the correlation length has an essential singularity as in Eq.(\ref{KT_sing}) \cite{Kosterlitz:74}. All quantities related to the correlation length exhibit the same KT essential singularity at the transition, for the $n$-OCP monolayer, KT-singularities are found for : the bond orientational order parameters $\phi_6$ (FIG.\ref{Fig4} (a)), the susceptibility $\chi_6$ (FIG.\ref{Fig4}(b) and FIG.\ref{Fig6}(d)) and correlations lengths $\xi(T)$ and $\xi_6(T)$ (FIG.\ref{Fig4}(c,d)). The transition temperatures $T_{F/H}$ or coupling constants $\Gamma_{F/H}$ found by fitting the KT-singularities of these quantities agree well and permit to locate the Hexatic/Fluid transition for several values of the power $n$ of inverse power law interaction of Eq.(\ref{IPLpot}). These results are reported on TABLE \ref{Table1}. As shown also in the previous section, the Kosterlitz-Thouless essential singularities can be very well studied with the multiple histogram method ; additional computations with different sizes and others values of $n$ are ongoing. These results for $n\leq 3$ are in excellent agreement with the KTNHY theory for the disclination-unbinding transition, they agree also with previous experimental \cite{Mehrotra:82,Gallet:82,Murray:87,Keim:07} and numerical \cite{He:03,Clark:09,Lin:06} studies.\\
In the present work, as $n$ is a continuous parameter in the IPL potential, we have been able to obtain empirical transition curves for both Solid/Hexatic and Hexatic/Fluid transition for $n\leq 3$. These curves are given by Eqs.(\ref{MeanField},\ref{Fit_gamma}) in Sect.III and they are represented on FIG.\ref{Fig1}(a) in the plane $(n,\rho)$. This phase diagram is extremely useful to locate the Hexatic phase for a given $n$ and to choose the range of numerical values for the surface density $\rho$, the temperature $T$ or the coupling constant $\Gamma$ in numerical computations that one has to use to observe the Hexatic phase, this is necessary to study in detail the properties of the Hexatic phase and of the Hexatic/Solid transition since the range in the coupling constant $\Gamma$ for which one can observe an Hexatic phase is very thin for almost all values of $n\leq3$. It is the reason for which, we consider the results obtained in this work as preliminary results for a longer study of the Hexatic phase in the $n$-OCP monolayers. Obviously, the accuracy of the transition curves reported in Sect.III, and the $\Gamma_i$ parameters given in TABLE \ref{Table1}, will be improved by adding data obtained with other values of $n$.\\ 
For $n > 3$, we found that the transition curves cross at $n\sim 4$, making $\rho_c^{F/H} > \rho_c^{H/S}$, and computations done at constant surface density and temperature (Runs A and B - FIG.\ref{Fig1}) indicate that the melting would not follow the KTHNY theory for $n>4$. The transition curves given in Sect.III do not permit to reproduce to a correct  accuracy  the data obtained in Runs A and B for $n\simeq 12.7$ and $n\simeq 7.5$. However, the finite size analysis of the fourth order cumulants using subsystems and some double peaked histograms of the bond orientational order parameters observed for $n=12.75-12.95$ may support the assumption that the melting becomes a first order transition. Nevertheless, we are not yet able to reach a definitive conclusion about the nature of the melting for $n\geq 4$ and more computations are needed for these values of $n$.\\
The recent results on the hard disks systems described by E.P. Bernard and W. Krauth \cite{Bernard:11}, show that for $n\rightarrow\infty$ the melting would be described by a first order Fluid/Hexatic transition followed by a second order Hexatic/Solid transition. These results do not agree neither with the KTHNY theory, nor with the grain-boundary induced melting proposed by Chui. However, these results might find a nice interpretation with the help of the theory of the melting based on isostructural critical points \cite{Bladon:95,Marcus:96,Chou:96} in which a first order transition between the Fluid and Hexatic phases is plausible \cite{Chou:96} ; this theory has also to be taken into account for studies with large but finite $n$.\\
The empirical transition curves for $n\leq 3$ and the excellent qualitative and semi-quantitative agreements with the KTHNY theory found for the melting of $n$-OCP monolayers with $n\leq 3$ are the main results of the present work. 
\acknowledgments
I am very grateful to Dominique Levesque, Jean-Michel Caillol, Jean-Jacques Weis, Emmanuel Trizac, Gerhard Kahl and Moritz Antlanger for many interesting discussions. The author acknowledges also the computation facilities (iDataPlex - IBM) provided by  {\it Direction Informatique\/} of {\it Universit\'e Paris-Sud\/}.\\
\appendix
\section{Ewald sums for Inverse Power Law interactions.}
In this appendix, we give the Ewald formulas used to compute the energy of the monolayer. A complete derivation of the Ewald method starting from the interaction potential of Eq.(\ref{IPLpot}) and for any continuous values of $n$ can be found in refs.\cite{Mazars:10,Mazars:11}.\\ 
For $n\neq2$, the energy of a configuration of the $n$-OCP monolayer is given by 
\begin{widetext}
\begin{equation}
\label{Ewald_1}
\hspace{-0.8in}\begin{array}{ll}
\displaystyle E &\displaystyle=\frac{Q^2}{2}\sum_{i=1}^{N}\sum_{j=1}^{N}\sum_{\mbox{\small $\bm{S}_{\bm{n}}$}}\mbox{}^{\prime}\frac{\Gamma\left(\frac{n}{2};\alpha^2 \mid\bm{s}_{ij}+\bm{S}_{\bm{n}}\mid^2 \right)}{\Gamma(\frac{n}{2}) \mid\bm{s}_{ij}+\bm{S}_{\bm{n}}\mid^n}+\frac{Q^2}{2}\frac{\pi}{A}\sum_{\bm{G}\neq 0} \left(\frac{G}{2}\right)^{(n-2)} \frac{\Gamma\left(1-\frac{n}{2}; \frac{G^2}{4\alpha^2}\right)}{\Gamma(\frac{n}{2})}\left|\sum_{i=1}^N \exp\left(j\bm{G}.\bm{s}_i\right)\right|^2\\
&\\
&\displaystyle-\frac{\alpha^{n}}{n\Gamma\left(\frac{n}{2}\right)}NQ^2-\frac{\pi}{A}\frac{\alpha^{(n-2)}}{(2-n)\Gamma\left(\frac{n}{2}\right)}N^2Q^2
\end{array}
\end{equation}
\end{widetext}
with $\bm{s}_i$ the location of the point particle in the monolayer, $\bm{S}_{\bm{n}}$ a symbolic notation for the periodic images of the system, $\bm{G}$ are the wave vectors of the reciprocal lattice associated to the periodic lattice $\bm{S}_{\bm{n}}$ and $\alpha$ is the Ewald parameter \cite{Mazars:11}. For $n=2$, the energy is given by 
\begin{equation}
\label{Ewald_3}
\begin{array}{ll}
\displaystyle E &\displaystyle=\frac{Q^2}{2}\sum_{i=1}^{N}\sum_{j=1}^{N}\sum_{\mbox{\small $\bm{S}_{\bm{n}}$}}\mbox{}^{\prime}\frac{\exp\left(-\alpha^2 \mid\bm{s}_{ij}+\bm{S}_{\bm{n}}\mid^2 \right)}{\mid\bm{s}_{ij}+\bm{S}_{\bm{n}}\mid^n}\\
&\\
&\displaystyle +\frac{Q^2}{2}\frac{\pi}{A}\sum_{\bm{G}\neq 0} \mbox{ E}_1\left(\frac{G^2}{4\alpha^2}\right)\left|\sum_{i=1}^N \exp\left(j\bm{G}.\bm{s}_i\right)\right|^2\\
&\\
&\displaystyle-\frac{\alpha^{2}}{2}NQ^2+\frac{\pi}{A}\ln(\alpha) N^2Q^2
\end{array}
\end{equation}
with $E_1(x)=-E_i(-x)$ the exponential integral.\\
In Eq.(\ref{Ewald_1}), $\Gamma(a ; x)$ is the complementary incomplete gamma function, when $a<0$ (i.e. $n>2$) the function has to be computed by using the recursive relation 
\begin{equation}
\label{Ewald_2}
\Gamma(a +1; x)=a\Gamma(a ; x)+x^a e^{-x}
\end{equation}


\vspace{2.in}

\begin{thebibliography}{0}
\bibitem{Mermin:68}  N.D. Mermin,  Phys. Rev., {\bf 176}, 250 (1968).
\bibitem{Jancovici:67}  B. Jancovici, Phys. Rev. Lett., {\bf 19}, 20, (1967).
\bibitem{Monarkha:book:03}  Y. Monarkha and K. Kono, {\it Two-Dimensional Coulomb Liquids and Solids - Springer Series in Solid-State Science\/} {\bf 142} (Springer, Berlin - 2004).
\bibitem{Grimes:79}  C.C. Grimes and G. Adams,  Phys. Rev. Lett., {\bf 42}, 795,(1979).
\bibitem{Mehrotra:82} R. Mehrotra, B.M. Guenin and A.J. Dahm, Phys. Rev. Lett., {\bf 48}, 641,(1982).
\bibitem{Gallet:82} F. Gallet, G. Deville, A. Vald\`es and F.I.B. Williams, Phys. Rev. Lett., {\bf 49}, 212,(1982).
\bibitem{Platzman:74}  P.M. Platzman and H. Fukuyama,  Phys. Rev. B, {\bf 10}, 3150 (1974).
\bibitem{Muto:99} S. Muto and H. Aoki, Phys. Rev. B, {\bf 59}, 14911 (1999).
\bibitem{He:03}  W.J. He, T.  Cui, Y.M. Ma, Z.M. Liu and G.T. Zou, Phys. Rev. B, {\bf 68}, 195104 (2003).
\bibitem{Clark:09}  B.K. Clark, M. Casula and D.M. Ceperley  Phys. Rev. Lett., {\bf 103}, 055701 (2009).
\bibitem{Nosenko:09}  V. Nosenko,  S.K. Zhdanov, A.V. Ivlev, C.A. Knapek and G.E. Morfill, Phys. Rev. Lett., {\bf 103}, 015001(2009).
\bibitem{Murray:87} C.A. Murray and D.H. Van Winkle,  Phys. Rev. Lett., {\bf 58}, 1200,(1987).
\bibitem{Keim:07} P. Keim, G. Maret, and H.H. von Gr\"unberg, Phys. Rev. E, {\bf 75}, 031402 (2007) ; P. Keim, G. Maret, U. Herz and H.H. von Gr\"unberg,   Phys. Rev. Lett., {\bf 92}, 215504 (2004).
\bibitem{Marcus:97}  A.H. Marcus and S.A. Rice, Phys. Rev. E, {\bf 55}, 637 (1997).
\bibitem{Dullens:04}  R.P.A. Dullens and W.K. Kegel,  Phys. Rev. Lett., {\bf 92}, 195702 (2004).
\bibitem{Zheng:06}  X.H. Zheng and R. Grieve, Phys. Rev. B, {\bf 73}, 064205 (2006).
\bibitem{Peng:11}  Y. Peng, Z. Wang, A.M. Alsayed, A.G. Yodh and Y. Han, Phys. Rev. Lett., {\bf 104}, 205703 (2010) ;  Phys. Rev. E, {\bf 83}, 011404, (2011).
\bibitem{Gribova:11}  N. Gribova, A. Arnold, T. Schilling and C. Holm, J. Chem. Phys., {\bf 135}, 054514, (2011).
\bibitem{Iaconis:10} J. Iaconis, R.G. Melko and A.A. Burkov,  Phys. Rev. B, {\bf 82}, 180504(R) (2010).
\bibitem{Saiki:11}  T. Saiki and R. Ikeda,  Phys. Rev. B, {\bf 83}, 174501 (2011).
\bibitem{Hadzibabic:06}  Z. Hadzibabic, P. Kr\"{u}ger, M. Cheneau, B. Battelier and J. Dalibard, Nature, {\bf 441}, 1118 (2006).
\bibitem{Choi:12} J.-Y. Choi, S. W. Seo and Y. Shin, cond-mat.quant-gas, arXiv:1211.5649 (2012). 
\bibitem{Negulyaev:09}  N.N. Negulyaev, {\it et al\/}  Phys. Rev. Lett., {\bf 102}, 246102 (2009).
\bibitem{Strandburg:88}  K.J. Strandburg, Rev. Mod. Phys., {\bf 60}, 161 (1988).
\bibitem{Chui:82}  S.T. Chui,  Phys. Rev. Lett., {\bf 48}, 933 (1982) ;   Phys. Rev. B, {\bf 28}, 178 (1983).
\bibitem{Nelson:book:83}  D.R. Nelson, {\it Defect-mediated Phase Transitions - Phase Transitions and Critical Phenomena\/} {\bf 7}, edited by C. Domb, and J.L. Lebowitz (Academic Press, London - 1983).
\bibitem{Nelson:78}  D.R. Nelson, Phys. Rev. B, {\bf 18}, 2318 (1978).
\bibitem{Nelson:79}  D.R. Nelson and B.I. Halperin, Phys. Rev. B, {\bf 19}, 2457 (1979).
\bibitem{Young:79} A.P. Young, Phys. Rev. B, {\bf 19}, 1855 (1979).
\bibitem{Zippelius:80} A.  Zippelius, B.I. Halperin and D.R. Nelson, Phys. Rev. B, {\bf 22}, 2514 (1980).
\bibitem{Kosterlitz:74}  J.M. Kosterlitz, J. Phys. C: Solid State Phys., {\bf 7}, 1046 (1974). 
\bibitem{Saito:82} Y. Saito, Phys. Rev. Lett., {\bf 48}, 1114 (1982) ; Phys. Rev. B, {\bf 26}, 6239 (1982) ; Phys. Rev. B, {\bf 27}, 6973 (1983).
\bibitem{Saito:83} Y. Saito, Surf. Sci., {\bf 125}, 285 (1983) ; H. M\"{u}ller-Krumbhaar and Y. Saito, Surf. Sci., {\bf 144}, 84 (1984).
\bibitem{Bladon:95} P. Bladon and D. Frenkel, Phys. Rev. Lett., {\bf 74}, 2519 (1995).
\bibitem{Marcus:96} A.H. Marcus and S.A. Rice, Phys. Rev. Lett., {\bf 77}, 2577 (1995) ; Phys. Rev. E, {\bf 55}, 637 (1997).
\bibitem{Chou:96} T. Chou and D.R. Nelson, Phys. Rev. E, {\bf 53}, 2560 (1996).
\bibitem{Prestipino:11}  S. Prestipino, F. Saija and P.V. Giaquinta, Phys. Rev. Lett., {\bf 106}, 235701 (2011).
\bibitem{Bernard:11}  E.P. Bernard and W. Krauth,  Phys. Rev. Lett., {\bf 107}, 155704 (2011).
\bibitem{Jaster:99}  A. Jaster, Phys. Rev. E, {\bf 59}, 2594 (1999).
\bibitem{Sengupta:00}  S. Sengupta, P. Nielaba and K. Binder Phys. Rev. E, {\bf 61}, 6294 (2000).
\bibitem{Pronk:04}  S. Pronk and D. Frenkel, Phys. Rev. E, {\bf 69}, 066123 (2004).
\bibitem{Mak:06} C.H. Mak,  Phys. Rev. E, {\bf 73}, 065104(R) (2006).
\bibitem{Wierschem:11}  K. Wierschem and E. Manousakis,  Phys. Rev. B, {\bf 83}, 214108 (2011).
\bibitem{Tobochnik:82} J. Tobochnik and G.V. Chester, Phys. Rev. B, {\bf 25}, 6778 (1982).
\bibitem{Shiba:09}  H. Shiba, H. Onuki and T. Araki, EPL, {\bf 86}, 66004 (2009).
\bibitem{Asenjo:11} D. Asenjo, F. Lund, S. Poblete, R. Soto and M. Sotomayor, Phys. Rev. B, {\bf 83}, 174110 (2011).
\bibitem{Isobe:10} M. Isobe and B.J. Alder, Prog. Theor. Phys. Suppl., {\bf 184}, 439 (2010).
\bibitem{Isobe:11} M. Isobe and B.J. Alder, J. Chem. Phys., {\bf 137}, 194501 (2012).
\bibitem{Mazars:11}  M. Mazars, Phys. Rep., {\bf 500}, 43 (2011).  
\bibitem{Mazars:10}  M. Mazars, J. Phys. A : Math. Theor., {\bf 43}, 425002 (2010).
\bibitem{Hoover:71}  W.G. Hoover, S.G. Gray and K.W. Johnson, J. Chem. Phys., {\bf 55}, 1128 (1971) ; W.G. Hoover, D.A. Young and R. Grover, J. Chem. Phys., {\bf 56}, 2207 (1972) .
\bibitem{Khrapak:11} S.A. Khrapak, M. Chaudhuri and G.E. Morfill, J. Chem. Phys., {\bf 134}, 241101 (2011) ; S.A. Khrapak and G.E. Morfill, Phys. Rev. Lett., {\bf 103}, 255003 (2009).
\bibitem{Groh:01}  B. Groh and S. Dietrich, Phys. Rev. E, {\bf 63}, 021203 (2001).
\bibitem{Pedersen:10} U.R. Pedersen, T.B. Schr{\o}der and J.C. Dyre, Phys. Rev. Lett., {\bf 105}, 157801 (2010) ; U.R. Pedersen, N.P. Bailey, T.B. Schr{\o}der and J.C. Dyre, Phys. Rev. Lett., {\bf 100}, 015701 (2008).
\bibitem{Coslovich:08} D. Coslovich and C.M. Roland, J. Phys. Chem. B, {\bf 112}, 1329 (2008).
\bibitem{Agrawal:95} R. Agrawal and D.A. Kofke, Phys. Rev. Lett., {\bf 74}, 122 (1995).
\bibitem{Lidmar:97} J. Lidmar and M. Wallin, Phys. Rev. B, {\bf 55}, 522 (1997).
\bibitem{Lin:06} S.Z. Lin, B. Zheng and S. Trimper,  Phys. Rev. E, {\bf 73}, 066106 (2006).
\bibitem{Michele:04} C. De Michele, F. Sciortino and A. Coniglio, J. Phys. : Condens. Matter, {\bf 16}, L489 (2004). 
\bibitem{Berthier:10} L. Berthier and G. Tarjus, Phys. Rev. E, {\bf 82}, 031502 (2010).
\bibitem{Casalini:06} R. Casalini, U. Mohanty and C.M. Roland, J. Chem. Phys., {\bf 125}, 014505 (2006).
\bibitem{Frenkel:12} D. Frenkel, {\it Simulations: the dark side.\/} Proceedings of International School of Physics "Enrico Fermi" Course CLXXXIV - {\it Physics of Complex Colloids\/}, arXiv:1211.4440 (2012).
\bibitem{Weber:95}  H. Weber, D. Marx and K. Binder, Phys. Rev. B, {\bf 51}, 14636 (1995). 
\bibitem{Privman:book:90} V. Privman, {\it Finite Size Scaling and Numerical Simulation of Statistical Systems\/}(World Scientific, Singapore - 1990).
\bibitem{Ferrenberg:88} A.M. Ferrenberg and R.H. Swendsen, Phys. Rev. Lett., {\bf 61}, 2635 (1988) ; Phys. Rev. Lett., {\bf 63}, 1195 (1989).
\bibitem{Newman:book:99} M.E.J. Newman and G.T. Barkema, {\it Monte Carlo Methods in Statistical Physics.\/} (Oxford University Press - 1999).
\bibitem{Mazars:08}  M. Mazars, EPL, {\bf 84}, 55002 (2008).
\bibitem{Weis:01}  J.-J. Weis, D. Levesque and S. Jorge,  Phys. Rev. B, {\bf 63}, 045308 (2001).
\bibitem{Totsuji:78} H. Totsuji, Phys. Rev. A, {\bf 17}, 399 (1978).
\bibitem{Smith:08}  E.R. Smith, J. Chem. Phys., {\bf 128}, 174104 (2008).
\bibitem{Fisher:72}  M.E. Fisher, S.-K. Ma and B.G. Nickel, Phys. Rev. Lett., {\bf 29}, 917 (1972).
\bibitem{Luijten:02}  E. Luijten and H.W.J. Bl\"{o}te, Phys. Rev. Lett., {\bf 89}, 025703 (2002). 
\end{thebibliography}
\end{document}